\documentclass[12pt]{article}

\usepackage[toc,page]{appendix}
\usepackage{slashed}
\usepackage[sumlimits,intlimits,namelimits]{amsmath} 
\usepackage[english]{babel}
\usepackage{graphicx}
\usepackage{hyperref}
\usepackage{cite}

\newcommand{\ba}{\begin{eqnarray}}
\newcommand{\ea}{\end{eqnarray}}

\newcommand{\ice}[1]{\relax}

\usepackage{color}
\usepackage{amsmath}
\usepackage{booktabs} 
\usepackage{subfigure}

\DeclareMathOperator{\Tr}{Tr}

\def\Li{\mbox{Li}}

\begin{document}

\begin{titlepage}

\begin{flushright}
SI-HEP-2021-29\\
SFB-257-P3H-21-084
\end{flushright}
\vspace{1.2cm}
\begin{center}
	  { \Large\bf NLO QCD Corrections to \\[2mm] Inclusive \boldmath $b \to c \ell \bar{\nu}$ \unboldmath Decay Spectra
	    up to \boldmath $1/m_Q^3$ \unboldmath  }
\end{center}
\vspace{0.5cm}
\begin{center}
{\sc Thomas Mannel}, {\sc Daniel Moreno} and {\sc Alexei A. Pivovarov} \\[0.2cm]
{\sf Center for Particle Physics Siegen, Theoretische Physik 1, Universit\"at Siegen\\ 57068 Siegen, Germany}
\end{center}

\vspace{0.8cm}
\begin{abstract}\noindent
  We present analytical results for higher order corrections to the decay spectra of
  inclusive semileptonic heavy hadron weak decays, using the heavy quark expansion (HQE).  
  We compute analytically the spectrum of the leptonic invariant mass 
  for $B \to X_c \ell \bar{\nu}$ up to and including terms of order $1/m_Q^3$ within the HQE 
  at next-to-leading order (NLO) in $\alpha_s$.
The full dependence of the differential rate on the mass of the final-state quark is 
taken into account.  
We discuss the implications of our results for the precision determination 
of the CKM matrix element $|V_{cb}|$.  
\end{abstract}

\end{titlepage}

\section{Introduction} 
\label{sec:Intro}
Testing the flavour sector of the Standard Model (SM) is one of the major current activities 
in particle physics. On the experimental side large experimental activities are under way, which 
have started with the construction of $B$ factories about two decades ago, and efforts are 
continuing with the BelleII experiment at the KEK Super-Flavour Factory in Tsukuba (Japan)
and at the LHCb experiment at the Large Hadron Collider (LHC) at CERN in Geneva (Switzerland). 

From the theoretical side, enormous progress has been made over the last three decades. Making 
use of the fact that the $b$-quark mass (and to some extent also the $c$ quark mass) are large 
compared to the QCD scale $\Lambda_{\rm QCD}$, precision methods have been developed, which allow 
us to perform a combined expansion in $\Lambda_{\rm QCD}/m_Q$ and $\alpha_s (m_Q)$~\cite{Shifman:1987rj,Eichten:1989zv,Isgur:1989vq,Grinstein:1990mj} 
resulting in precision predictions with controllable uncertainties.  
     
All this has established the flavour structure of the SM, namely the Cabbibo-Kobayashi-Maskawa
(CKM) picture of quark mixing at the precision level, leaving only limited room for physics 
beyond the SM (BSM). However, some recent data show persistent tensions with the SM predictions, 
the so-called $B$ anomalies, which could be interpreted as first signals for BSM effects. To this 
end, precision quark-flavour physics may become an important tool to establish the presence of 
BSM effects. However, this requires - aside from more precise measurements - very precise theoretical 
calculations, which require in the field of flavour physics also to deal with non-perturbative 
effects. In the context of the heavy quark expansion (HQE) this means to push for higher orders in both $\Lambda_{\rm QCD}/m_Q$ 
and $\alpha_s (m_Q)$.        

One important example is the calculation of the differential rate for the inclusive 
$B \to X_c \ell \bar{\nu}$ within the HQE\cite{Chay:1990da,Bigi:1992su,Blok:1993va,Manohar:1993qn}, which is the key ingredient for the precision determination
of $|V_{cb}|$ from inclusive decays, where a theoretical uncertainty of less than two percent has been 
achieved. In fact, the tension between the values of $|V_{xb}|$ ($x=u, c$) 
extracted from inclusive and exclusive decays is one of the persistent $b$ anomalies. 

The HQE hadronic parameters constitute the non-perturbative input into the HQE. The number of independent 
parameters strongly proliferates at higher orders, thereby limiting our possibilities for a fully 
model independent determination of $|V_{cb}|$ from inclusive decays. Up to $(\Lambda_{\rm QCD}/m_Q)^3$ 
only four HQE parameters emerge, which can be extracted from the measurement of moments of the 
charged lepton energy spectrum and the hadronic invariant mass spectrum, but starting at 
$(\Lambda_{\rm QCD}/m_Q)^4$ the number of independent HQE parameters is too large to be extracted from data. 

However, it has been shown in~\cite{Mannel:2018mqv,Fael:2018vsp} that one may exploit a symmetry of the 
HQE, the so-called Reparametrization Invariance (RPI), which allows one to reduce the number of independent 
HQE parameters for specific observables. For the case at hand, the inclusive semileptonic 
$B \to X_c \ell \bar{\nu}$ decays, these observables are all related to the spectrum of the leptonic 
invariant mass. Based on this approach a further improvement of the precision of the inclusive 
$|V_{cb}|$ determination is expected since the reduced number of HQE parameters can be determined 
from the data.  

The current status of the HQE calculation of $B \to X_c \ell \bar{\nu}$ is already quite elaborate. 
The leading term, i.e. the partonic rate is known at 
N$^2$LO-QCD~\cite{Czarnecki:1989bz,Gremm:1996gg,Falk:1997jq,Falk:1995me,Trott:2004xc,Aquila:2005hq,Czarnecki:1997hc,Czarnecki:1998kt,Melnikov:2008qs,Pak:2008qt,Pak:2008cp,Dowling:2008ap,Biswas:2009rb,Gambino:2011cq} 
and at N$^3$LO only for the total rate~\cite{Fael:2020tow}. 
The first power correction, of order $(\Lambda_{\rm QCD}/m_Q)^2$, is 
known at NLO-QCD~\cite{Bigi:1992su,Blok:1993va,Manohar:1993qn,Becher:2007tk,Alberti:2012dn,Alberti:2013kxa,Mannel:2014xza,Mannel:2015wsa,Mannel:2015jka}. 
The second power correction, of order $(\Lambda_{\rm QCD}/m_Q)^3$, is known at LO-QCD~\cite{Gremm:1996df,Mannel:2017jfk} 
and at NLO-QCD only for the total width~\cite{Mannel:2019qel}. Finally, third and fourth power corrections, of order $(\Lambda_{\rm QCD}/m_Q)^{4,5}$ have 
been computed at LO-QCD in \cite{Bigi:2009ym,Mannel:2010wj}.

In the present paper we give an analytical result 
for the leptonic invariant mass spectrum at order $\alpha_s(m_Q) (\Lambda_{\rm QCD}/m_Q)^3$. 
We also give the analytical result for the $\alpha_s(m_Q) (\Lambda_{\rm QCD}/m_Q)^2$ terms of the 
leptonic invariant mass spectrum, which are known up to now only numerically~\cite{Alberti:2013kxa}. 
These results will allow us to extract $|V_{cb}|$ from the RPI method on the basis of analytical expressions.

We provide an ancillary file called ``Coef.m''  which contains analytical results in Mathematica format for the 
coefficients of the leptonic invariant mass spectrum up to $\mathcal{O}(\alpha_s(m_Q)/m_Q^3)$ and the $\mathcal{O}(\alpha_s(m_Q))$ Darwin coefficients
of the moments with a low cut in the leptonic invariant mass. 
It can be downloaded in arXiv by using the link in ``ancillary files'' or by downloading the entire source package as a 
gzipped tar file (.tar.gz).

The paper is organized as follows. In Sec.~\ref{sec:rate} and Sec.~\ref{sec:HQE} we set the notation and briefly describe our 
method for the computation of the differential width. In Sec.~\ref{difratemb2} and Sec.~\ref{difratemb3} we compute the 
$(\Lambda_{\rm QCD}/m_Q)^2$ and $(\Lambda_{\rm QCD}/m_Q)^3$ HQE coefficients of the leptonic invariant mass spectrum
at $\mathcal{O}(\alpha_s(m_Q))$. Finally, we discuss the impact of our results after a brief numerical analysis in Sec.~\ref{sec:disc}. 
Analytical results for the coefficients are displayed in the appendix.

\section{HQE for Inclusive Heavy Flavour Decays}
\label{sec:rate}
We start giving some 
basics of the theoretical description of inclusive semileptonic
decays within the HQE. More details can be found in the literature (e.g.~\cite{Bigi:1993fe}).
The effective Fermi Lagrangian ${\cal{L}}_{\rm eff}$
for the semileptonic $b \to c \ell \bar{\nu}_\ell$ transitions reads 
\begin{equation}
  \label{eq:FermiLagr}
{\cal L}_{\rm eff} = 2\sqrt{2}G_F V_{cb}(\bar{b}_L \gamma_\mu c_L) 
(\bar{\nu}_L \gamma^\mu \ell_L) + {\rm h.c.} \, , 
\end{equation}
with the subscript $L$ denoting the left-handed 
fermion fields. Here $G_F$ is the Fermi constant 
and $V_{cb}$ is the relevant CKM matrix element. 

Using the optical theorem one obtains the inclusive decay rate
$B\to X_c\ell\bar{\nu}_\ell$ from taking an absorptive part of 
the forward matrix element of the leading order transition operator
${\cal T}$
\begin{equation}\label{eq:trans_operator}
{\cal T} = i\!\!\int \! dx\,    
T\left\{ {\cal L}_{\rm eff} (x)  {\cal L}_{\rm eff} (0) \right\} \, ,
\quad \Gamma (B \to X_c \ell \bar{\nu}_\ell)
\sim \text{Im} \langle B|{\cal T} |B\rangle .
\end{equation} 
Since the heavy quark mass $m_b$ is a large scale compared to the hadronization 
scale $\Lambda_{\rm QCD}$ of QCD ($m_Q\gg\Lambda_{\rm QCD}$), the 
forward matrix element contains perturbatively calculable 
contributions. These can be separated from the non-perturbative
pieces using the method of effective field theory. 

For a heavy hadron with momentum $p_B$ and mass $M_B$, a large part
of the heavy-quark momentum $p_b$ is due to a pure kinematical contribution due to 
its large mass $p_b=m_b v+\Delta$ with $v=p_B/M_B$ being the velocity of
the heavy hadron. The momentum $\Delta \sim {\cal O} (\Lambda_{\rm QCD})$ describes 
the soft-scale fluctuations of the heavy quark field near its mass shell.
This decomposition of the quark momentum is implemented 
by re-defining the heavy quark field $b(x)$ 
\begin{equation}\label{eq:heavy-quark-me-phase}
b(x) = e^{-i m_b(v x)}b_v (x)\, .
\end{equation}
so that $\partial b_v (x) \sim \Delta$. 
Inserting this into (\ref{eq:trans_operator}) we get 
\begin{equation}\label{eq:trans_operator1}
{\cal T} = i\!\!\int \! dx\,    e^{i m_b v \cdot x} 
T\left\{ \widetilde{\cal L}_{\rm eff} (x)  \widetilde{\cal L}_{\rm eff} (0) \right\} \, ,
\end{equation}
where $\widetilde{\cal L}$ is the same expression as ${\cal L}$ with the replacement 
$b(x) \to b_v(x) $. 

This expression allows one to set up the HQE as     
an expansion in $\Lambda_{\rm QCD}/m_b$ by matching the imaginary part of the 
transition operator ${\cal T}$ in QCD
to an expansion in inverse powers of the heavy quark mass using local operators defined in 
Heavy Quark Effective Theory (HQET)~\cite{Mannel:1991mc,Manohar:1997qy} with matching coefficients $C_i$
\begin{eqnarray}
	\label{eq:HQE-1}
	\mbox{Im}\, \mathcal{T} = \Gamma^0 |V_{cb}|^2 
	\bigg( C_0 \mathcal{O}_0 
	+ C_v \frac{\mathcal{O}_v}{m_b} 
	+ C_\pi \frac{\mathcal{O}_\pi}{2m_b^2} 
	+ C_G \frac{\mathcal{O}_G}{2m_b^2} 
	+ C_D \frac{\mathcal{O}_D}{4m_b^3}
	+ C_{LS} \frac{\mathcal{O}_{LS}}{4m_b^3}
	\bigg)\,.
\end{eqnarray}
The HQET operators $\mathcal{O}_i$ are listed below ordered by their mass dimension up to dimension six
\begin{eqnarray}
 \mathcal{O}_0 &=& \bar h_v h_v \qquad\qquad\qquad\qquad\qquad\qquad\;\; \mbox{(mass dimension three)}\,,
 \\
 \mathcal{O}_v &=& \bar h_v v\cdot \pi h_v \qquad\qquad\qquad\qquad\qquad\;\;\; \mbox{(mass dimension four)}\,,
 \\
 \mathcal{O}_\pi &=& \bar h_v \pi_\perp^2 h_v \qquad\qquad\qquad\qquad\qquad\quad\;\; \mbox{(mass dimension five)}\,,
 \label{mupi} \\
 \mathcal{O}_G &=& \frac{1}{2}\bar h_v [\gamma^\mu, \gamma^\nu] \pi_{\perp\,\mu}\pi_{\perp\,\nu}  h_v \quad\quad\qquad\quad\;\; \mbox{(mass dimension five) }\,,
 \label{muG} \\
 \mathcal{O}_D &=& \bar h_v[\pi_{\perp\,\mu},[\pi_{\perp}^\mu , v\cdot \pi]] h_v \qquad\qquad\qquad \mbox{(mass dimension six)}\,,
 \label{rhoD}\\
 \mathcal{O}_{LS} &=& \frac{1}{2}\bar h_v[\gamma^\mu,\gamma^\nu]\{ \pi_{\perp\,\mu},[\pi_{\perp\,\nu}, v\cdot \pi] \} h_v \quad\; \mbox{(mass dimension six)}\,, 
 \label{Ops} \label{rhoLS}
\end{eqnarray}
where $\pi_\mu = i D_\mu = i\partial_\mu +g_s A_\mu^a T^a$ is the covariant derivative of QCD,
$\pi^\mu =v^\mu (v\pi)+\pi^\mu_\perp$ and where we have neglected operators which are of higher dimension on shell. 
Here the field $h_v$ is the HQET field, whose dynamics is determined by the HQET Lagrangian~\cite{Manohar:1997qy}.

It is convenient to choose the local operator $\bar b \slashed v b$ defined in full QCD as the 
leading term of the HQE in Eq.~(\ref{eq:HQE-1}) instead of $\mathcal{O}_0$, 
since its forward matrix element with hadronic states is absolutely
normalized. The HQE of the operator  $\bar b \slashed v b$ reads
\begin{equation}
\bar b \slashed v b = \mathcal{O}_0 + \tilde{C}_v \frac{\mathcal{O}_v}{m_b} + \tilde C_\pi \frac{\mathcal{O}_\pi}{2m_b^2} + \tilde C_G \frac{\mathcal{O}_G}{2m_b^2} 
 + \tilde C_D \frac{\mathcal{O}_D}{4m_b^3}
 + \tilde C_{LS} \frac{\mathcal{O}_{LS} }{4m_b^3}\,,
 \label{hqebvb}
\end{equation}
with the matching coefficients $\tilde{C}_i$ being pure numbers.
Eventually we use the equations of motion (EOM) of the HQET 
Lagrangian to get rid of the operator $\mathcal{O}_v$ in Eq.~(\ref{eq:HQE-1}).

Thus, the HQE for semileptonic weak decays
is written as~(e.g.~\cite{Benson:2003kp})
\begin{equation}
\label{hqewidth}
 \Gamma(B\rightarrow X_c \ell \bar \nu_\ell ) = \Gamma^0 |V_{cb}|^2 
 \bigg[   C_0 
- C_{\mu_\pi}\frac{\mu_\pi^2}{2m_b^2}
+ C_{\mu_G}\frac{\mu_G^2}{2m_b^2}
- C_{\rho_D} \frac{\rho_D^3}{2m_b^3}
 - C_{\rho_{LS}} \frac{\rho_{LS}^3}{2m_b^3}
 \bigg]\,,
\end{equation}
where $\Gamma^0= G_F^2 m_b^5/(192 \pi^3)$ and $m_b$ is the $b$-quark mass. 
The coefficients $C_i$, $i=0,\mu_\pi,\mu_G,\rho_D,\rho_{LS}$ depend 
(in case of neglecting the lepton and light-quark masses) on the ratio $\rho= m_c^2/m_b^2$, where $m_c$ is the $c$-quark mass. Note that from 
reparametrization invariance $C_0 = C_{\mu_\pi}$ and $C_{\mu_G}=C_{\rho_{LS}}$~\cite{Mannel:2018mqv,Manohar:2010sf}.

The parameters $\mu_\pi^2$, $\mu_G^2$, $\rho_D^3$ and $\rho_{LS}^3$ are forward matrix elements 
of local operators usually called the hadronic
parameters of the HQE. The definition of these parameters in
our calculation reads 
\begin{eqnarray}
 \langle B(p_B)\lvert \bar b \slashed v b \lvert B(p_B)\rangle &=& 2M_B\,,  \\
 - \langle B(p_B)\lvert \mathcal{O}_\pi \lvert B(p_B)\rangle &=& 2M_B \mu_\pi^2\,, \\ 
 c_F(\mu)\langle B(p_B)\lvert \mathcal{O}_G \lvert B(p_B)\rangle
 &=& 2M_B \mu_G^2\,, \\
  - c_D(\mu)\langle B(p_B)\lvert \mathcal{O}_D \lvert B(p_B)\rangle&=& 4M_B \rho_D^3\,, \\
 -  c_S(\mu)\langle B(p_B)\lvert \mathcal{O}_{LS} \lvert B(p_B)\rangle&=& 4 M_B \rho_{LS}^3\,.
\end{eqnarray}
where the forward matrix elements are taken over the physical state
of the heavy meson or, theoretically, in full QCD~\cite{Mannel:2018mqv}.
The quantities $c_F(\mu)$, $c_D(\mu)$, and $c_S(\mu)$ are matching
coefficients in the HQET Lagrangian with $\mu$ being the renormalization
point. The NLO expressions for these coefficients are known. 
The expression given in Eq.~(\ref{hqewidth}) emerges
from the direct matching of 
the QCD expression for the transition operator to HQET. 
Taking the forward matrix element of Eq.~(\ref{eq:HQE-1}) after using the HQE of the $\bar b \slashed v b$ operator and the EOM of the 
HQET Lagrangian yields Eq.~(\ref{hqewidth}).

In general, there is an additional operator 
${\cal O}_{\rm{I}} = \bar{h}_v (v\pi)^2 h_v $ in the complete basis of dimension 
five operators, however, it will be of higher order in the HQE
after using the EOM of HQET. Similarly, there are five
additional operators at dimension six which vanish (or become of
higher order in the power expansion) 
after using the EOM.

The coefficients $C_i$ have a perturbative expansion in the 
strong coupling constant $\alpha_s (m_b)$. The leading coefficient $C_0$ 
is known analytically to~${\cal O}(\alpha_s^2)$ 
precision in the massless limit for the final 
state quark~\cite{vanRitbergen:1999gs}.
At this order, the mass corrections have been analytically accounted 
for the total width as an expansion in the mass of the final
fermion in~\cite{Pak:2008qt} and for the differential distribution
in~\cite{Melnikov:2008qs}. For the total rate, ${\cal O}(\alpha_s^3)$ corrections have been computed quite recently~\cite{Fael:2020tow}
with mass corrections of the final state quark again accounted as an expansion.

The coefficient of the kinetic energy parameter is linked to the coefficient 
$C_0$ by reparametrization invariance (for an explicit check see,
e.g.~\cite{Becher:2007tk}). 
The NLO correction to the coefficient of the chromo-magnetic
parameter $C_{\mu_G^2}$
has been investigated in~\cite{Alberti:2013kxa} 
where the hadronic tensor has been computed analytically and the total decay rate has been then obtained by direct numerical
integration over the phase space.
This calculation allows for the
application of different energy/momentum cuts in the phase space
necessary for the accurate comparison with experimental data. 

\section{Lepton Invariant-Mass Spectrum: Generalities}
\label{sec:HQE}
In this section we will discuss the setup for the calculation of the differential rate in the leptonic 
invariant mass squared $r = q^2/m_b^2$. We follow the general
procedure described
in~\cite{Mannel:2021ubk}, and use
the dispersion representation of the one-loop diagram to 
write the lepton-neutrino loop 
as an integral differential in the lepton pair invariant mass squared.
In this way the leptonic loop 
becomes an  ``effective massive propagator'' of mass $q$. 

We assume the leptons to be massles, which leads to 
\begin{equation}
\label{relrhosrho1LS}
\int\frac{d^D k}{(2\pi)^D}\frac{-\Tr(\Gamma^\sigma i(\slashed k + \slashed q_2)\Gamma^\rho i\slashed k)}{k^2(k+q_2)^2}  
  = i\int_{0}^{\infty}d(q^2)  \frac{\rho_s(q^2)}{q^2 - q_2^2-i\eta}(q_2^2 g^{\rho\sigma}-q_2^\rho q_2^\sigma)\,,
  \quad \Gamma_\mu  = \gamma_\mu \frac{1}{2}(1-\gamma_5)\,,
\end{equation}
where $D = 4 -2 \epsilon$, $q_2$ is the four-momentum flowing through the leptons, 
and the spectral density $\rho_s$ to $\mathcal{O}(\epsilon^0)$ is a constant
\begin{equation}
 \rho_s (q^2) =  
 \frac{2}{3}\frac{1}{16\pi^2} + \mathcal{O}(\epsilon)\,.
\end{equation}
Since renormalization can be performed at the differential level, the integrand is finite, and it is enough to keep 
the $\mathcal{O}(\epsilon^0)$ term in $\rho_s$. 
 
The purely leptonic part is not affected by QCD corrections,
which keeps the spectral density very simple. 
Note that the ``effective massive propagator'' of mass $q$ is transverse due to the fact that the leptons are massless. 
After writing the leptonic loop in this form, we can compute the width differential 
in the dilepton pair invariant mass squared, which we write as it follows
\begin{eqnarray}
 \label{hqedifwidth}
\frac{d\Gamma(B\rightarrow X_c \ell \bar \nu_\ell )}{dr}
 &=& \Gamma^0 |V_{cb}|^2 
 \rho_s
 \bigg[   \mathcal{C}_0 
- \mathcal{C}_{\mu_\pi}\frac{\mu_\pi^2}{2m_b^2}
+ \mathcal{C}_{\mu_G}\frac{\mu_G^2}{2m_b^2}
- \mathcal{C}_{\rho_D} \frac{\rho_D^3}{2m_b^3}
 - \mathcal{C}_{\rho_{LS}} \frac{\rho_{LS}^3}{2m_b^3}
 \bigg]
  \nonumber
  \\
  &=& \Gamma^0 |V_{cb}|^2 
  \rho_s
 \bigg[ \mathcal{C}_0 \bigg( 1 
- \frac{\bar{\mathcal{C}}_\pi - \bar{\mathcal{C}}_v }{\mathcal{C}_0}\frac{\mu_\pi^2}{2m_b^2}\bigg)
+ \bigg(\frac{\bar{\mathcal{C}}_G}{c_F(\mu)} -  \bar{\mathcal{C}}_v \bigg)\frac{\mu_G^2}{2m_b^2}
\nonumber
\\
&&
- \bigg(\frac{\bar{\mathcal{C}}_D}{c_D(\mu)}-\frac{1}{2} \bar{\mathcal{C}}_v \bigg) \frac{\rho_D^3}{2m_b^3}
 - \bigg(\frac{\bar{\mathcal{C}}_{LS} }{c_S(\mu)} - \frac{1}{2} \bar{\mathcal{C}}_v \bigg) \frac{\rho_{LS}^3}{2m_b^3}
 \bigg]\, , 
\end{eqnarray}
where $0 \le r \le (1-\sqrt{\rho})^2$ and we have defined $\bar{\mathcal{C}}_i\equiv \mathcal{C}_i - \mathcal{C}_0 \tilde{C}_i$ as the difference 
between the coefficients $\mathcal{C}_i$ of the HQE of the transition operator (in differential form) and the current multiplied by $\mathcal{C}_0$.  
The coefficients $\mathcal{C}_i$ of the differential rate depend on
two variables $r$ and $\rho$ and are related to the coefficients of the total rate
by
\begin{eqnarray}
 C_i(\rho) &=& \int_{0}^{(1-\sqrt{\rho})^2}dr\,\rho_s \, \mathcal{C}_i(r,\rho)\,.
\end{eqnarray}
As discussed in ref.~\cite{Fael:2018vsp} one can
reduce the number of independent HQE parameters 
by using reparametrization-invariant observables, and the lepton invariant-mass spectrum is such 
a quantity. However, in the same way as for the lepton energy moments the resulting expression 
cannot be interpreted point-by-point, and so one refers to moments of the spectra, for which a 
clean HQE exists. In addition, it may be necessary from the experimental 
side to introduce cuts on the variables, for which the leptonic invariant mass still preserves RPI.

Consequently we discuss the  
(not yet normalized) moments with a lower cut $r_{\rm min}$ and upper
cut $r_{\rm max}$ that are defined as 
\begin{eqnarray}
  M_n (\rho,r_{\mbox{\scriptsize min}},r_{\mbox{\scriptsize max}}) &=& 
  \int_{r_{\mbox{\scriptsize min}}}^{r_{\mbox{\scriptsize max}}}dr\, r^n \frac{d\Gamma(r,\rho)}{dr} \,,
\end{eqnarray}
with $0<r_{\mbox{\scriptsize min}}<r_{\mbox{\scriptsize max}}<(1-\sqrt{\rho})^2$. Since we derive 
analytical expressions for the leptonic
invariant-mass spectrum the above quantity can be readily evaluated 
numerically, except for the Darwin coefficient, which requires
keeping $D=4-2\epsilon$ in order to regularize the IR divergence that appears at the 
end of the integration region for $r$.
 
\begin{figure}[!htb]  
	\centering
	\includegraphics[width=1.0\textwidth]{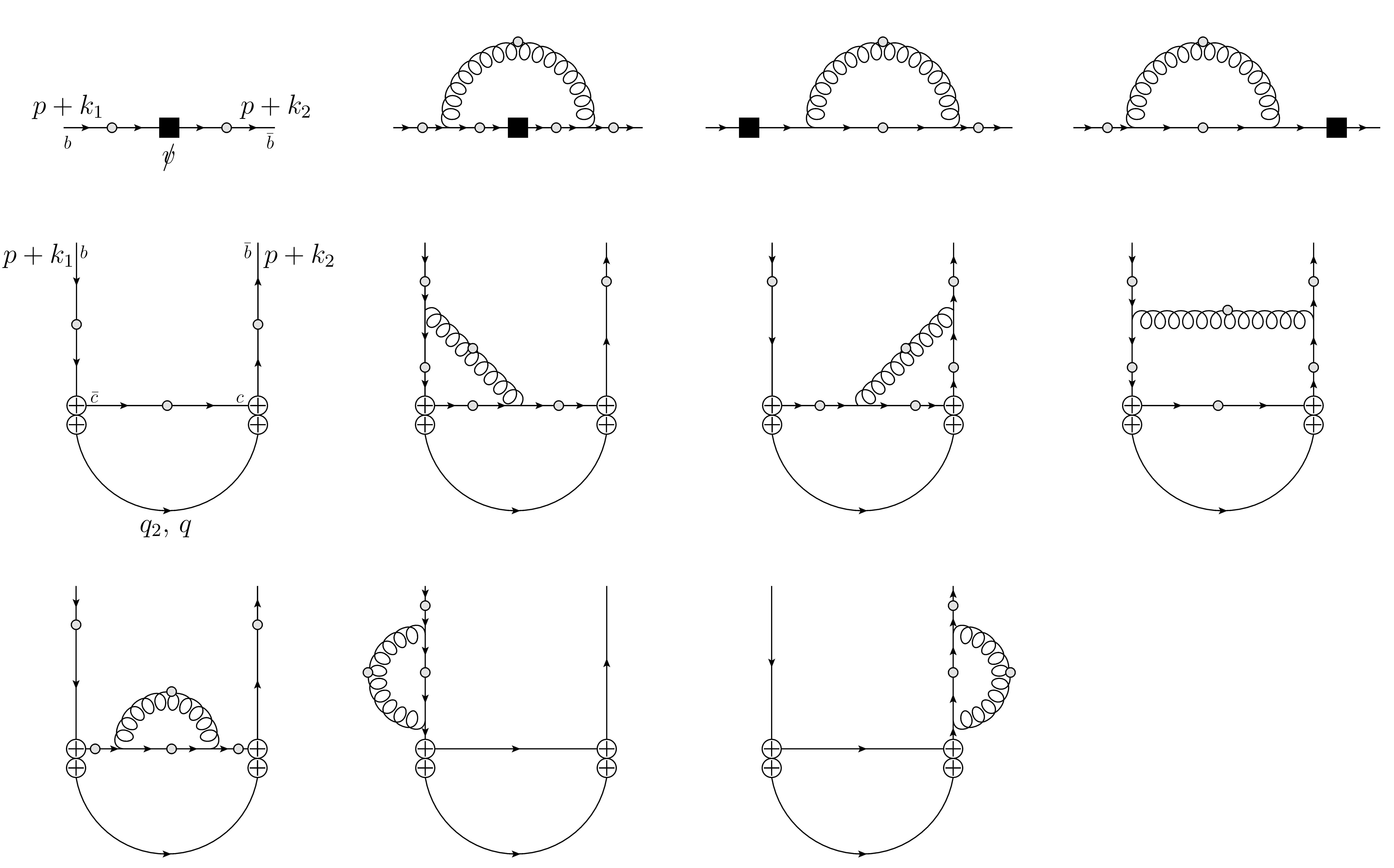}
        \caption{Quark to quark-gluon scattering diagrams contributing to the coefficients $\bar{\mathcal{C}}_i = \mathcal{C}_i - \mathcal{C}_0 \tilde{C}_i$ 
        of power corrections in the HQE of the $b \to c \ell \bar{\nu}$ decay spectrum, Eq.~(\ref{hqedifwidth}). Black squares stand for $\slashed v$ insertions, 
        circles with crosses for insertions of the operator in $\mathcal{L}_{\rm eff}$ and grey dots stand for possible gluon insertions with incoming momentum $k_2-k_1$. 
        After properly accounting for all one gluon insertions, there are five diagrams at LO-QCD and fourty-one diagrams at NLO-QCD.
        }
        \label{SampleFDdifwidth}
\end{figure}

For the leading power, the Feynman diagrams contributing to the
differential
width at LO-QCD and NLO-QCD are 
one-loop and two-loop quark to quark scattering diagrams.
For the computation of power corrections, the LO-QCD and NLO-QCD contributions 
are represented by one-loop and two-loop quark to quark-gluon
scattering Feynman
diagrams, respectively. 
The latter are shown in Fig.~[\ref{SampleFDdifwidth}]. 

By using LiteRed~\cite{Lee:2012cn,Lee:2013mka} the corresponding amplitude is reduced to a combination of the master integrals 
computed in Ref.~\cite{Mannel:2021ubk}. We use standard dimensional
regularization 
in $D=4-2\epsilon$ spacetime dimensions. 
Algebraic manipulations including Dirac algebra are carried out 
with the help of Tracer~\cite{Jamin:1991dp}. 
Expansion of Hypergeometric functions is done with the 
help of HypExp~\cite{Huber:2005yg,Huber:2007dx}. 
The computation is done in the Feynman gauge and we use the background
field method
to compute the scattering in the external gluonic field. 

We adopt the $\overline{\mbox{MS}}$ 
renormalization scheme~(e.g.~\cite{Grozin:2005yg})
for the renormalization of the strong coupling $\alpha_s(\mu)$ and the HQET Lagrangian.
The bottom and charm quarks will be renormalized on-shell. In practice, that is $b_B = (Z_2^{\mbox{\scriptsize OS}})^{1/2} b$ and 
$m_{c,B} = Z_{m_c}^{\mbox{\scriptsize OS}} m_c$, where the subscript B
denotes bare quantities,
the ones with no subscript 
stand for renormalized, and
\begin{eqnarray}
 Z_{m_q}^{\mbox{\scriptsize OS}} &=& 1 - C_F \frac{\alpha_s(\mu)}{4\pi}\bigg( \frac{3}{\epsilon} + 6 \ln\left(\frac{\mu}{m_q}\right) + 4 \bigg)\,,
\end{eqnarray}
with $Z_2^{\mbox{\scriptsize OS}}= Z_{m_b}^{\mbox{\scriptsize OS}}$ to this order. 
Also $g_{s,B}^2 = 4\pi Z_g^2 \alpha_s(\mu)\bar{\mu}^{2\epsilon}$, 
where $\bar \mu^{2} = \mu^{2}(e^{\gamma_E}/4\pi)$ is the $\overline{\mbox{MS}}$ renormalization scale. 
For the precision of the calculation, the renormalization factor of the strong coupling is only needed at tree level ($Z_g=1$).
The $SU(3)$ color factors are $C_F=4/3$ and $C_A =3$. 

We quote our results in the on-shell scheme
for both quark masses $m_c$ and $m_b$. For more precise predictions one 
usually chooses for the bottom quark a low-scale short distance mass
such as the kinetic or the $1S$ mass, and thus one needs to convert 
the on-shell mass into such a mass scheme. However, the known one-loop expression will be sufficient.      
For the charm quark mass we are free to choose still a different renormalization scheme; one can either renormalize 
it on-shell or in the $\overline{\mbox{MS}}$-scheme. 
We have chosen the former because results become slightly more compact in that scheme, and thus quote our results in the 
on-shell scheme. A change in the scheme can easily be achieved by using the 
relation between the $\overline{\mbox{MS}}$ and pole masses at one-loop order
\begin{eqnarray}
 m_c^{\mbox{\scriptsize pole}} &=& 
 m_c^{\overline{\mbox{\scriptsize MS}}}(\mu)\bigg(1 + C_F \frac{\alpha_s}{4\pi}\bigg( 6\ln\left(\frac{\mu}{m_c}\right) + 4\bigg)\bigg)\,.
\end{eqnarray}

\section{Differential Rate in the Lepton Invariant Mass at \boldmath ${\cal O} (1/m_b^2)$ \unboldmath}
\label{difratemb2}
The NLO correction to the terms at order $1/m_b^2$ have been computed already a while 
ago~\cite{Becher:2007tk,Alberti:2012dn,Alberti:2013kxa,Mannel:2014xza,Mannel:2015wsa,Mannel:2015jka}, but to the 
best of our knowledge there is no analytical expression for the leptonic invaraint-mass spectrum in the 
literature. The coefficient of $\mu_\pi^2$ can be 
inferred from the leading order term in the HQE, whereas the coefficient of $\mu_G^2$ must be computed explicitly. 
For the computation of the chromomagnetic term we follow the approach used earlier for the 
calculation of the total width~\cite{Mannel:2014xza,Mannel:2015wsa,Mannel:2015jka}. 

One takes the amplitude of quark to quark-gluon scattering, expands to linear order in the small momentum and
projects it to the corresponding dimension five HQET operator. At dimension five the operators (\ref{mupi}) and 
(\ref{muG}) appear together with the operator 
$\mathcal{O}_{{\scriptsize\mbox{I}}} = \bar h_v (v\cdot\pi)^2 h_v $,  
which is irrelevant because it is of higher order due to the EOM. 

The contribution to the chromomagnetic 
coefficient is obtained by considering a single small gluon momentum
$k_\perp$ and picking up 
the contribution antisymmetric in 
$k_\perp \epsilon_\perp$, where $\epsilon$ is the gluon polarization vector. 
Incoming and outgoing bottom quarks carry momentum $p$ and $p+k_\perp$ respectively, with $p^2=m_b^2$.

First we directly compute the difference between the HQE of the transition operator and the current

\begin{eqnarray}
 \bar{\mathcal{C}}_G &\equiv& \mathcal{C}_G - \mathcal{C}_0 \tilde{C}_G
 = Z_2^{\mbox{\scriptsize ON}}Z_{\mathcal{O}_G}(\mathcal{C}_G^{\mbox{\scriptsize bare}}
 - \mathcal{C}_0^{\mbox{\scriptsize bare}}  \tilde{C}_G^{\mbox{\scriptsize bare}})\,,
\end{eqnarray}
where 

\begin{equation}
 Z_{\mathcal{O}_G} = 1 - C_A \frac{\alpha_s}{4\pi} \frac{1}{\epsilon}\,,
\end{equation}
is the renormalization constant of the chromomagnetic operator. The
quantity $\bar{\mathcal{C}}_G$ 
is finite. 
Once determined, the coefficient in front of the matrix element in the differential width $\mathcal{C}_{\mu_G}$ is given by

\begin{equation}
 \mathcal{C}_{\mu_G} = \frac{\bar{\mathcal{C}}_G}{c_F(\mu)} - \bar{\mathcal{C}}_v\,,
\end{equation}
where

\begin{equation}
 c_F(\mu) = 1 + \frac{\alpha_s}{2\pi}\bigg[ C_F + C_A\bigg(1 + \ln\left(\frac{\mu}{m_b}\right)\bigg) \bigg]\,,
\end{equation}
is the coefficient of the chromomagnetic operator in the HQET Lagrangian~\cite{Manohar:1997qy}.

We still have to discuss how to compute the coefficient $\bar{\mathcal{C}}_v$. In order to obtain it one takes the amplitude 
of quark to quark-gluon scattering, expands to zeroth order in the small momentum and
projects it to the corresponding HQET operator. The contribution to
the $\mathcal{O}_v$ coefficient is obtained by considering a longitudinally polarized gluon 
exchange ($v\cdot\epsilon$) without momentum transfer. 
Incoming and outgoing bottom quarks carry momentum $p$, with $p^2=m_b^2$. Like in the previous case, we directly compute the 
difference between the HQE of the transition operator and the current
\begin{eqnarray}
 \bar{\mathcal{C}}_v &\equiv& \mathcal{C}_v - \mathcal{C}_0 \tilde{C}_v
 = Z_2^{\mbox{\scriptsize ON}}(\mathcal{C}_v^{\mbox{\scriptsize bare}}
 - \mathcal{C}_0^{\mbox{\scriptsize bare}}  \tilde{C}_v^{\mbox{\scriptsize bare}})\,.
\end{eqnarray}
After integration over $r$ in its whole allowed range we obtain the
known results 
for the coefficients of the total width $C_v$ and $C_{\mu_G}$ obtained in 
Refs.~\cite{Mannel:2014xza,Mannel:2015wsa,Mannel:2015jka}. We take the occasion to correct a misprint in Eq.(84) of Ref.~\cite{Mannel:2015jka}. 
The last term should carry an overall minus sign instead of plus sign. 

The resulting expressions are somewhat lengthy and we give the analytical result in the appendix~\ref{App:coefdifW}.  
We also provide them as text file ``Coef.m'' in Mathematica format.

\section{Differential Rate in the Lepton Invariant Mass at \boldmath ${\cal O} (1/m_b^3)$ \unboldmath}
\label{difratemb3}
It has been shown in ref.~\cite{Mannel:2018mqv}
that due to RPI only the coefficient of $\rho_D^3$ needs to be 
determined at ${\cal O} (1/m_b^3)$, since the coefficient of $\rho_{\rm LS}^3$ can be inferred from the one of $\mu_G$. Thus we will 
get the full answer at NLO for reparametrization invariant quantities by computing the coefficient of the 
Darwin term. 

To determine this coefficient
one takes the amplitude of quark to quark-gluon scattering, expands to quadratic order in the small momenta and
projects it to the corresponding dimension six HQET operator. At dimension six the operators (\ref{rhoD}, \ref{rhoLS}) 
appear, together with the following operators 
\begin{eqnarray}
 && \mathcal{O}_{{\scriptsize\mbox{II}}} = \bar h_v  (v\cdot \pi)\pi_\perp^2 h_v \;,\quad\quad
 \mathcal{O}_{{\scriptsize\mbox{III}}} = \bar h_v \pi_\perp^2 (v\cdot\pi) h_v \;,\quad\quad
 \mathcal{O}_{{\scriptsize\mbox{IV}}} = \bar h_v (v\cdot \pi)^3 h_v \;,
 \nonumber
 \\
 &&
 \mathcal{O}_{{\scriptsize\mbox{V}}} = \bar h_v
 \frac{1}{2}[\gamma^\mu,\gamma^\nu] 
\pi_{\perp\,\mu} \pi_{\perp\,\nu} (v\cdot\pi)h_v \;,\quad\quad
 \mathcal{O}_{{\scriptsize\mbox{VI}}} = \bar h_v (v\cdot\pi)
 \frac{1}{2}[\gamma^\mu,\gamma^\nu] 
\pi_{\perp\,\mu} \pi_{\perp\,\nu} h_v\,,
 \nonumber
\end{eqnarray}
which mix with (\ref{rhoD},\ref{rhoLS}),
but they contribute only to higher orders due to the EOM.   
We disentangle the mixing of such operators with the Darwin term by
considering a gluon 
with longitudinal polarization ($v\cdot\epsilon$), 
using two quark momenta $k_1$ and $k_2$ and picking up the structure
$k_{1\,\perp} k_{2\,\perp}$ symmetric in $k_1$, $k_2$~\cite{Mannel:2019qel}. 
The kinematical configuration is such that the incoming and outgoing bottom quarks carry momentum $p+k_{1\,\perp}$ and $p+k_{2\,\perp}$ respectively, 
with $p^2=m_b^2$. The gluon carries momentum $k_{2\,\perp}-k_{1\,\perp}$.
Note that the antisymetric part gives the spin-orbit term.

The matching is performed by integrating out the charm quark simultaneously
with the hard modes of the $b$-quark. This means that we treat $m_c^2/m_b^2$ as a number   
of order one in the limit $m_b \to \infty$,
implying that also $m_c \to \infty$. 
Therefore our results cannot be extrapolated to the limit $m_c\to 0$. 

For calculations of radiative corrections we use dimensional
regularization.
At NLO the cancellation of poles provides a solid check of the
computation. At the $1/m_b^3$ order this cancellation is quite delicate
since it
requires to consider the mixing between operators of different
dimension, which is known to happen in
HQET~\cite{Falk:1990pz,Bauer:1997gs,Finkemeier:1996uu,Balzereit:1996yy,Blok:1996iz,Lee:1991hp,Moreno:2017sgd,Lobregat:2018tmn,Moreno:2018lbo}.

In analogy to Sec.~\ref{difratemb2} we directly
compute the difference between the HQE of the transition operator
itself and the current $\bar b \slashed{v} b$
\begin{eqnarray}
\bar{\mathcal{C}}_D \equiv \mathcal{C}_D - \mathcal{C}_0  \tilde{C}_D 
&=& Z_2^{\mbox{\scriptsize ON}}Z_{\mathcal{O}_D}(\mathcal{C}_D^{\mbox{\scriptsize bare}} - \mathcal{C}_0^{\mbox{\scriptsize bare}}  \tilde{C}_D^{\mbox{\scriptsize bare}}) 
+ \delta \bar{\mathcal{C}}_{D}^{mix}\,,
\end{eqnarray}
where 
\begin{eqnarray}
 Z_{\mathcal{O}_D} &=& - \frac{1}{6}C_A \frac{\alpha_s}{\pi}\frac{1}{\epsilon}\,,
 \\
\delta \bar{\mathcal{C}}_{D}^{mix} &=& 
  \bigg[
   C_F\bigg( 
     \frac{4}{3}\bar{\mathcal{C}}_\pi
   - \frac{2}{3}\bar{\mathcal{C}}_v
   \bigg) 
 + C_A\bigg(
  \frac{5}{12}\bar{\mathcal{C}}_G  
 + \frac{1}{12}\bar{\mathcal{C}}_\pi 
 - \frac{1}{4}\bar{\mathcal{C}}_v 
 \bigg)
 \bigg]
 \frac{\alpha_s}{\pi}\frac{1}{\epsilon}\,,
 \label{CDmix}
\end{eqnarray}
are the renormalization constant of the Darwin operator and
the contribution to the Darwin coefficient coming 
from the operator mixing in HQET, respectively. The quantity $\bar{\mathcal{C}}_D$ is finite. 

The corresponding anomalous dimensions in the operator mixing are
inferred from the cancellation 
of poles. Due to the functional structure of the coefficients
$\mathcal{C}_i$
the anomalous dimensions are determined 
uniquely. After taking into account the
combinatorial factors coming from the combined insertion of operators of the HQE and 
operators of the HQET Lagrangian we find that the anomalous
dimensions proportional to $\bar{\mathcal{C}}_D$, 
$\bar{\mathcal{C}}_G$, $\bar{\mathcal{C}}_\pi$ coincide with
known results~\cite{Bauer:1997gs}, which is a 
strong check of the calculation. The presence of the
coefficient $\bar{\mathcal{C}}_v$ means an admixture to the operator ${\cal O}_v$. 
Such an admixture was pointed out to be present in~\cite{Mannel:2019qel}.

Once $\bar{\mathcal{C}}_D$ is determined, the coefficient
in front of the matrix element of the differential width 
$\mathcal{C}_{\rho_D}$ is given by
\begin{equation}
 \mathcal{C}_{\rho_D} = \frac{\bar{\mathcal{C}}_D}{c_D(\mu)} - \frac{1}{2} \bar{\mathcal{C}}_v\,,
\end{equation}
where
\begin{equation}
 c_D(\mu) = 1 + \frac{\alpha_s}{\pi}\bigg[ 
 C_F\bigg(-\frac{8}{3}\ln\left(\frac{\mu}{m_b}\right) \bigg) 
 + C_A\bigg(\frac{1}{2} - \frac{2}{3}\ln\left(\frac{\mu}{m_b}\right)\bigg) \bigg]\,
\end{equation}
is the coefficient of the Darwin operator in the HQET Lagrangian~\cite{Manohar:1997qy}.

The analytical result for the coefficient ${\cal C}_{\rho_D} (r,\rho)$ is lengthy. 
We give it in the appendix~\ref{App:coefdifW} and supply it in the text file ``Coef.m''.

After integration over the lepton invariant mass $r$ in the allowed
range,
we obtain the coefficient of the total width $C_{\rho_D}$, which is displayed 
in Eqs.~(\ref{CrhoDtotLO}), (\ref{CrhoDCFtotNLO}) and \ref{CrhoDCAtotNLO}) in the appendix. 
Analytical expressions for moments with and without low $r$ cut are also computed and 
given in appendix \ref{App:Mnrcut} and \ref{App:Mn0}, respectively.
Expressions are very lengthy and are provided in the text file ``Coef.m''.

For the Darwin term, performing the last integration to obtain the coefficient of the total 
width or moments is rather subtle
since the coefficient is IR singular at the upper integration limit $r=(1-\sqrt{\rho})^2$, 
pointing out that expansion in $\epsilon$ and integration over $r$ do not commute. 
The coefficient of the total width and moments can be
obtained by restoring the $\epsilon$ dependence in the IR singular terms. 
The IR pole is $(-r + (1-\sqrt{\rho})^2)^{-3/2-\epsilon}$ whose integral over 
$r$ is finite in dimensional regularization due to the power $3/2$ and
there is no generation of new poles.
In other words, the formally divergent integrals are defined
in dimensional regularization that may look a bit uneasy if the
parameter $\epsilon$ is omitted
\begin{equation}
  \int_0^1 \frac{d\zeta}{(1-\zeta)^{3/2}}=
  \int_0^1
   \frac{d\zeta}{\zeta^{3/2}}=-2=B(-\frac{1}{2},1)=\frac{\Gamma(-\frac{1}{2})}{\Gamma(\frac{1}{2})}\,.
\end{equation}
The integral should be understood as a limit at $\epsilon\to 0$ of the
standard dimensionally regularized expression
\begin{equation}
  \int_0^1 \frac{d\zeta}{(1-\zeta)^{3/2+\epsilon}}\,.
\end{equation}
At LO our expression for the total width agrees with the known result~\cite{Gremm:1996df}. 
At NLO our result for the total width differs from the first calculation
of
ref.~\cite{Mannel:2019qel} based on the direct use of three loop
Feynman integrals
for the total decay rate.
The source of the disagreement is not yet clear 
but we consider the result of the present paper as the most reliable
one.
The reason for this claim is that the operator mixing inferred
in~\cite{Mannel:2019qel}
from the requirement of poles cancellation
differs from that emerging in the HQET Lagrangian.
In the present paper, we still infered some anomalous dimensions from
the explicit cancellation of the poles but part of the poles have been canceled with the
standard HQET operator mixing~\cite{Bauer:1997gs,Finkemeier:1996uu,Balzereit:1996yy,Blok:1996iz}.
We plan to investigate this problem further.

\begin{figure}[ht]
\centering
\subfigure[Leading power coefficient.]{\includegraphics[scale=0.66]{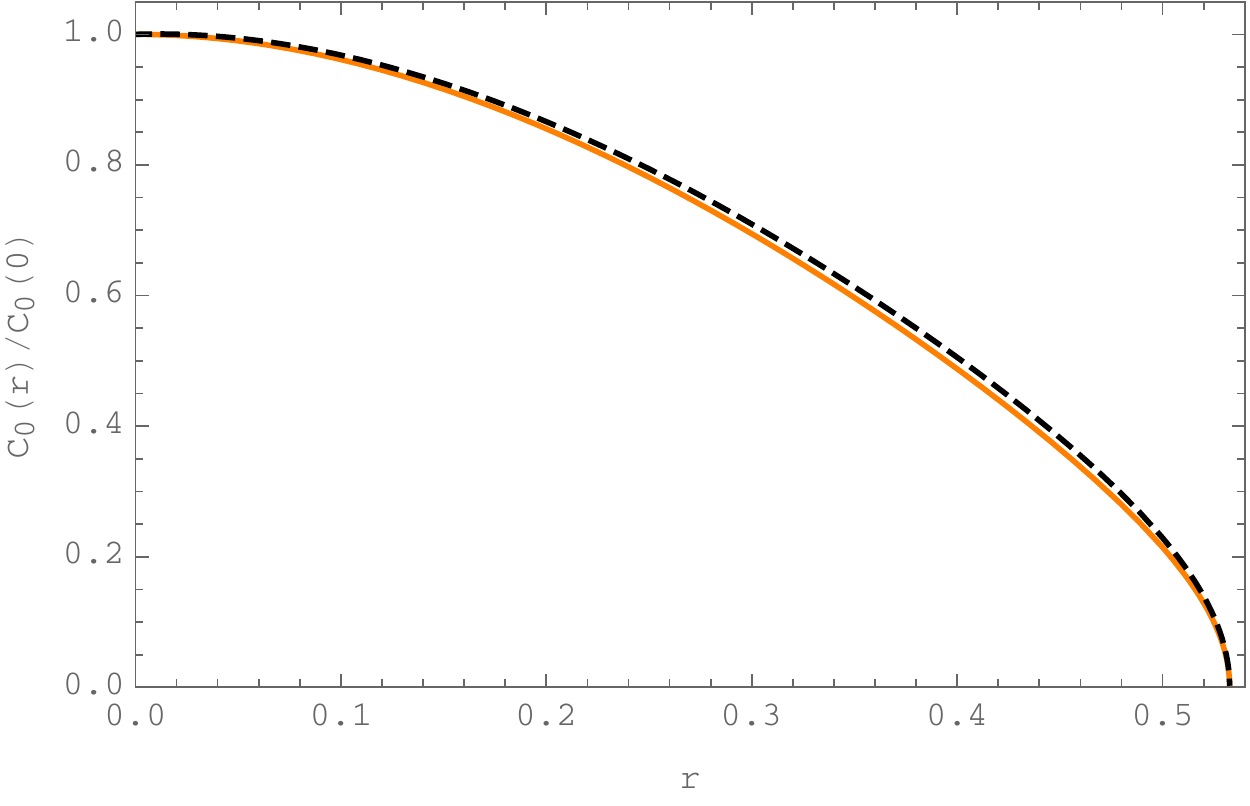}
\label{fig:subfigure1}}
\quad
\subfigure[Chromomagnetic operator coefficient.]{%
\includegraphics[scale=0.67]{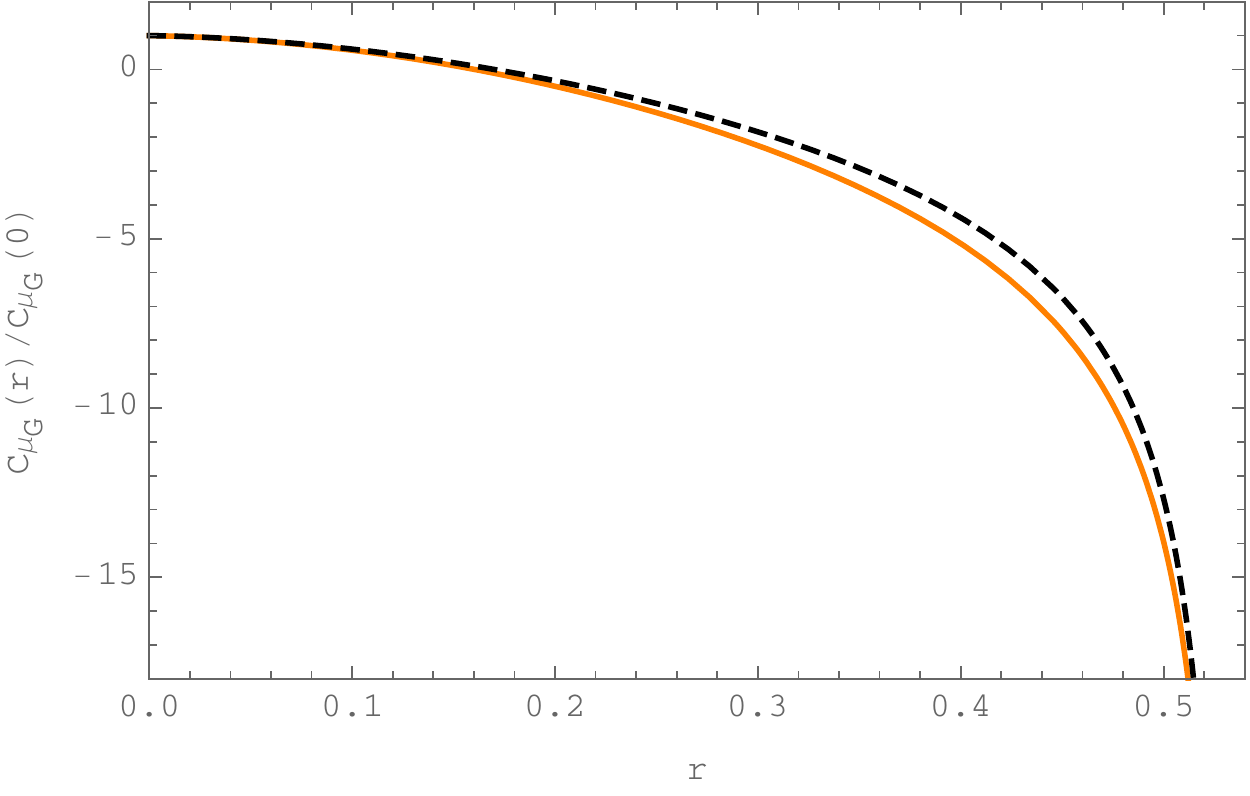}
\label{fig:subfigure3}}
\subfigure[Darwin operator coefficient.]{%
\includegraphics[scale=0.7]{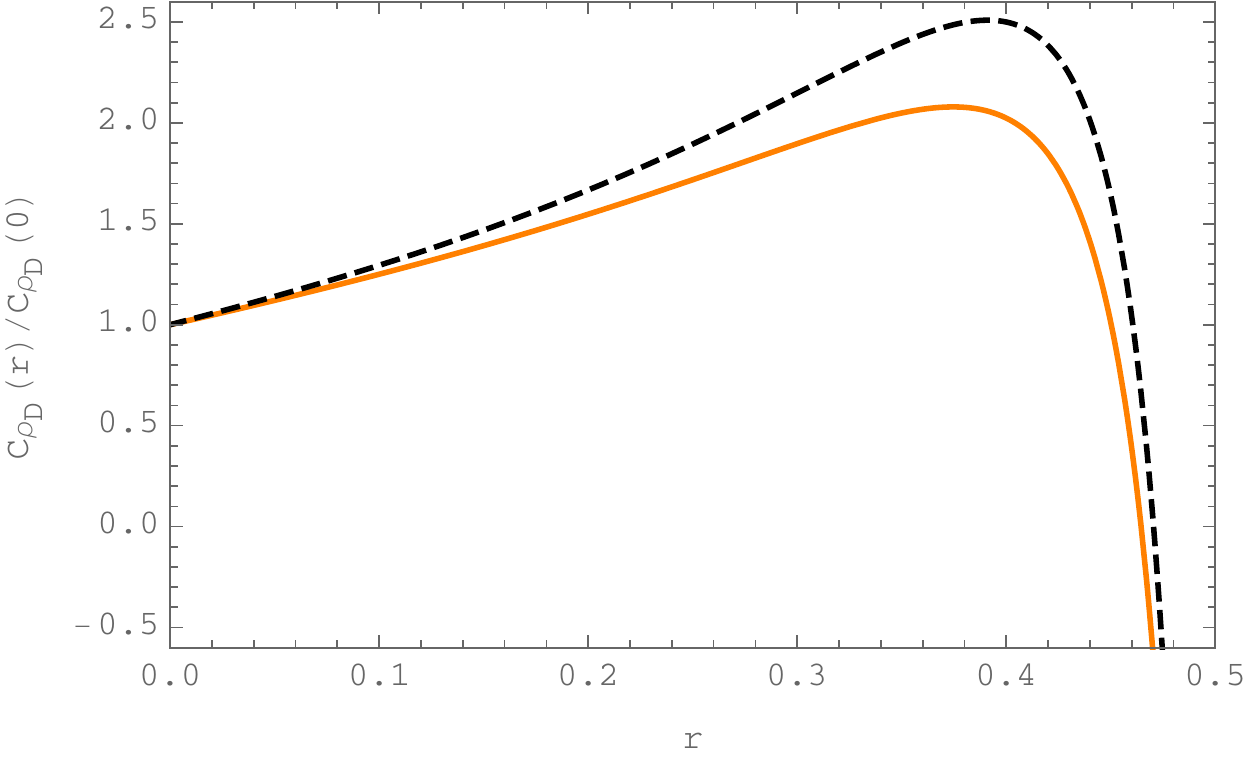}
\label{fig:subfigure2}}
\quad
\caption{Coefficients of the differential rate normalized to their value at $r=0$ as a function of the leptonic pair invariant mass squared $r$. 
The orange continuous line and the black dashed line stand for coefficients at LO and NLO, respectively.}
\label{fig:coefdGdq2}
\end{figure}

\begin{figure}[ht]
\centering
\subfigure[Plot in the allowed range $0<r<(1-\sqrt{\rho})^2$.]{\includegraphics[scale=0.66]{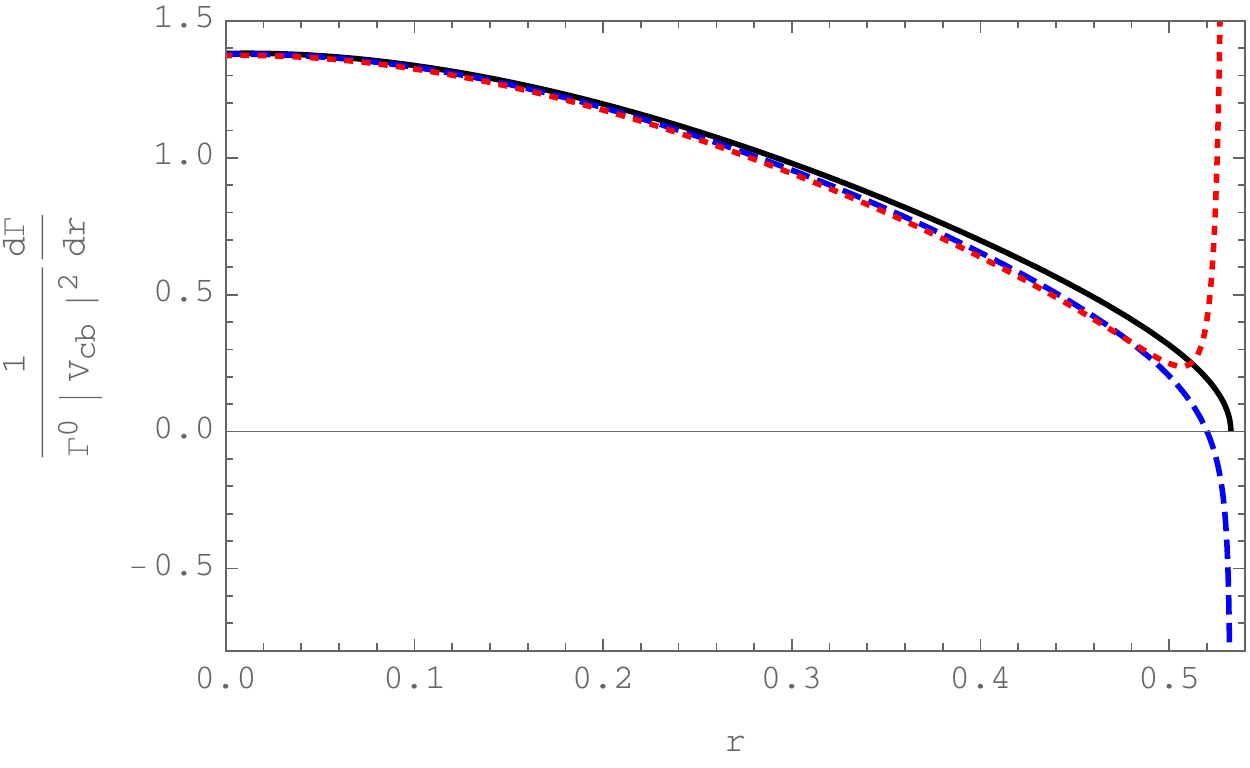}
\label{fig:dGdq2a}}
\quad
\subfigure[Plot in the range $0.2<r<0.3$.]{%
\includegraphics[scale=0.66]{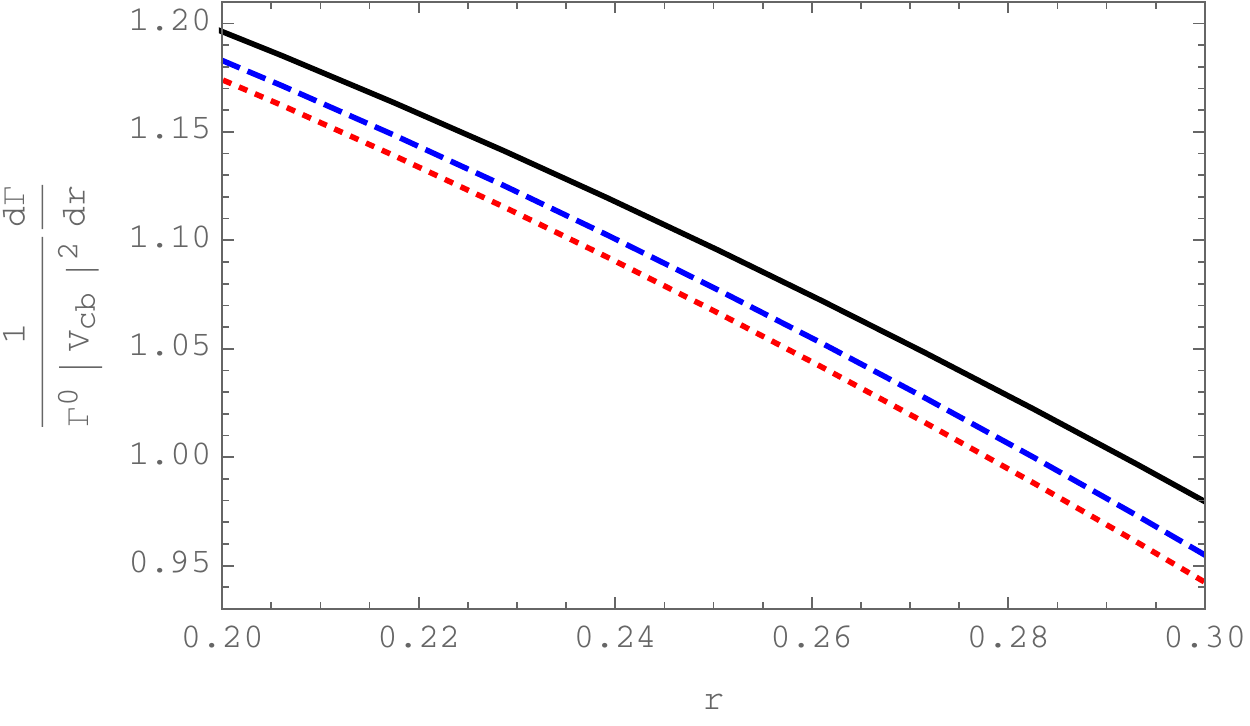}
\label{fig:dGdq2b}}
\caption{Differential rate as a function of the lepton pair invariant mass squared $r$ including subsequent power corrections at NLO.
The black continuous line stands for the leading power contribution, the blue dashed line includes $1/m_b^2$ corrections and the 
red dotted line includes $1/m_b^3$ corrections. 
The infrared singular behaviour at the end of the phase space becomes more abrupt for higher power corrections.}
\label{fig:dGdq2}
\end{figure}

To discuss the impact of our result, we show
in Figs.~[\ref{fig:coefdGdq2}] and [\ref{fig:dGdq2}] the dependence in the lepton 
pair invariant mass squared $r$ of the coefficients
of the differential rate normalized to their value at $r=0$, 
and of the differential rate by including different
corrections. We also plot the differential rate in a cut out range in order to perceive 
the size of corrections. For illustration, we use the typical
numerical values displayed in table~\ref{tab:par}. 
For the matrix elements these values are taken from~\cite{Benson:2003kp}.

\begin{table}
 \begin{center}
\begin{tabular}{|c|c|}
\hline
 Parameter & Numerical value\\ 
   \hline
 $\mu=m_b$ & $4.8$ GeV \\ 
 $\rho=m_c^2/m_b^2$& $0.073$ \\ 
 $\alpha_s(m_b)$ & $0.215$\\
 $\mu_\pi^2$ & $0.4$ GeV$^2$\\
 $\mu_G^2$ &  $0.35$ GeV$^2$\\ 
 $\rho_{D}^3$ & $0.2$ GeV$^3$\\ 
 $\rho_{LS}^3$ & $-0.15$ GeV$^3$\\
 $r_{\rm max}=(1-\sqrt{\rho})^2$ & $0.5326$\\
 $q^2_{\rm max}= m_b^2 r_{\rm max} $ & $12.27$ GeV$^2$\\
 \hline
\end{tabular}
\caption{Numerical values of parameters used in plots.\label{tab:par}}
\end{center}
\end{table}

\section{Numerical analysis}
\label{sec:disc}
With this paper, the full expression for the leptonic invariant-mass spectrum (including also cut moments in the leptonic invariant-mass squared) 
in $b \to c \ell \bar{\nu}$ decays is known analytically up to the order $\alpha_s/m_b^3$ with massive final state quark. This will allow one
to increase the precision 
of the $|V_{cb}|$ determination compared to previous analyses by using leptonic invariant-mass
moments with cuts in this variable.
The RPI of HQE is useful for the inclusion of terms 
of order $1/m_b^4$, where the numerical values of the HQE parameters at
this order 
can be extracted
from data~\cite{Fael:2018vsp}. 

Experimentally, one measures the moments of the spectrum.
The low $q^2$ is difficult to detect and the experimentalist use cuts
while integrating all over the remaining phase space up to the available $q^2$.
The normalization of the moments onto themselves gives a great deal of
simplification experimentally because requires then only counting events without
absolute normalization of the rate. Therefore, the
theoretical predictions are written in the form of
normalized moments with different cuts

\begin{equation}
\langle q^{2n}\rangle\equiv m_b^{2n}\hat{M}_n(r_{\rm cut}) = m_b^{2n}M_n(r_{\rm cut})/M_0(r_{\rm cut})
\end{equation}
The experimental data set is available with cuts in $q^2$
from $q_{\rm cut}^2 = 3.0~{\rm GeV^2}$ to $q_{\rm cut}^2=10.0~{\rm
  GeV^2}$
in steps of $0.5~{\rm GeV^2}$~\cite{Belle:2021idw}.

\begin{figure}[ht]
\centering
\subfigure[Plot in the range $q^2_{\rm cut}<10.5$ GeV$^2$.]{\includegraphics[scale=0.66]{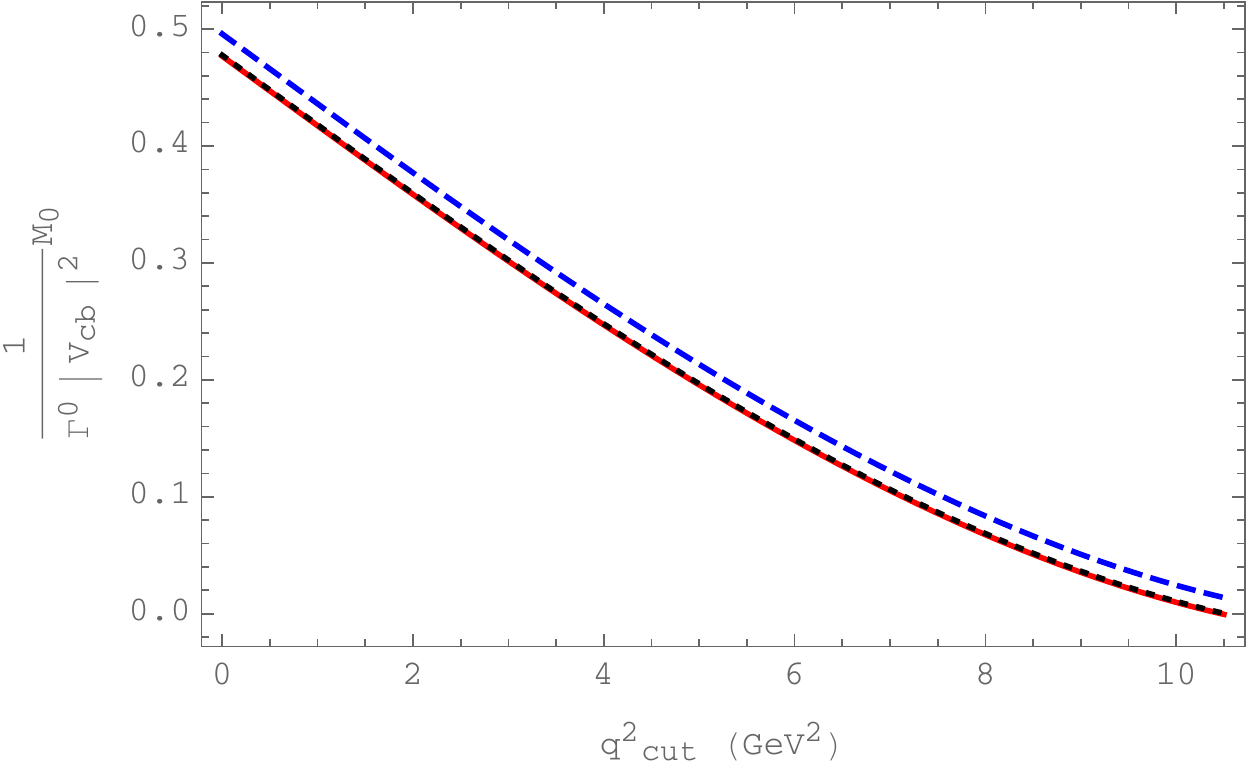}
\label{fig:M0a}}
\quad
\subfigure[Plot in the range $4.5\mbox{ GeV}^2<q^2_{\rm cut}<5$ GeV$^2$.]{%
\includegraphics[scale=0.66]{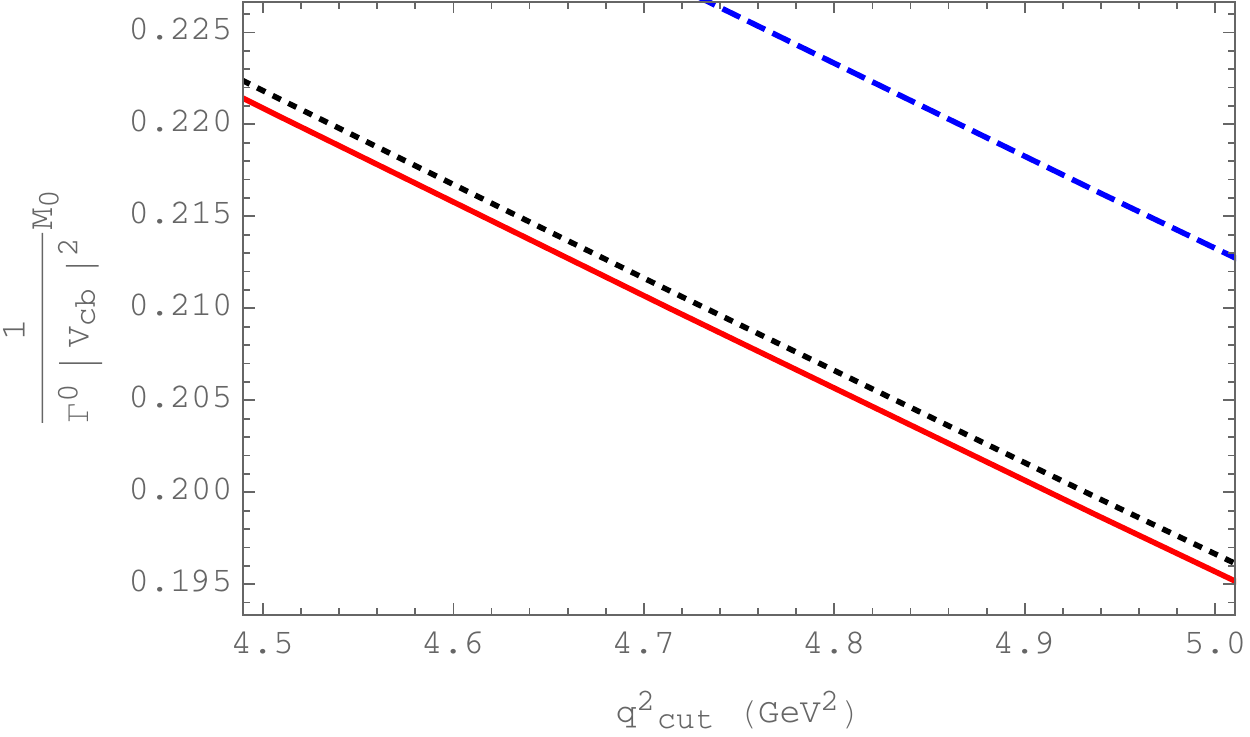}
\label{fig:M0b}}
\caption{Partial rate (zeroth moment) as a function of the low cut $q^2_{\rm cut}$ in the lepton pair invariant mass squared.
The blue dashed line includes NLO corrections up to $1/m_b^2$, the black dotted line includes also $1/m_b^3$ corrections at LO, and 
the continuous red line includes NLO corrections up to $1/m_b^3$.}
\label{fig:M0}
\end{figure}

\begin{figure}[ht]
\centering
\subfigure[First moment.]{\includegraphics[scale=0.65]{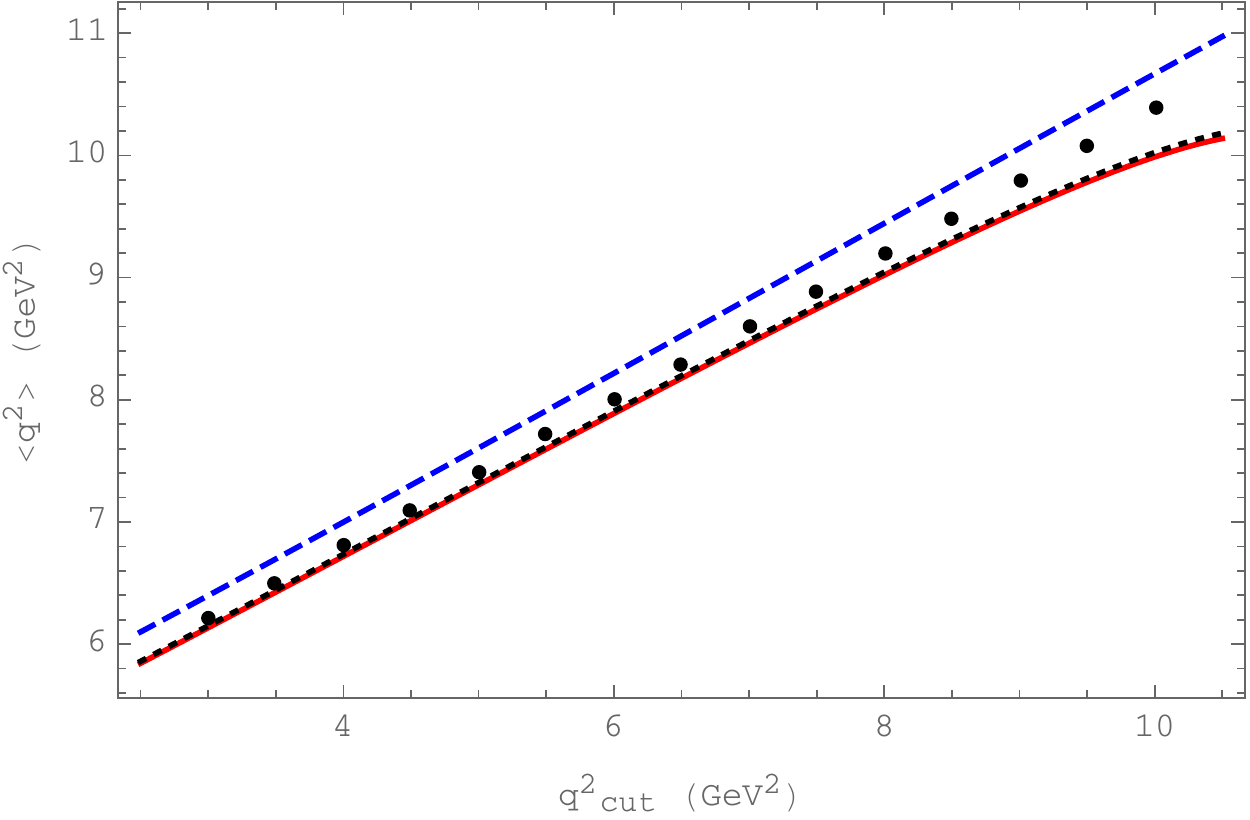}
\label{fig:subfigure1}}
\quad
\subfigure[Second moment.]{%
\includegraphics[scale=0.65]{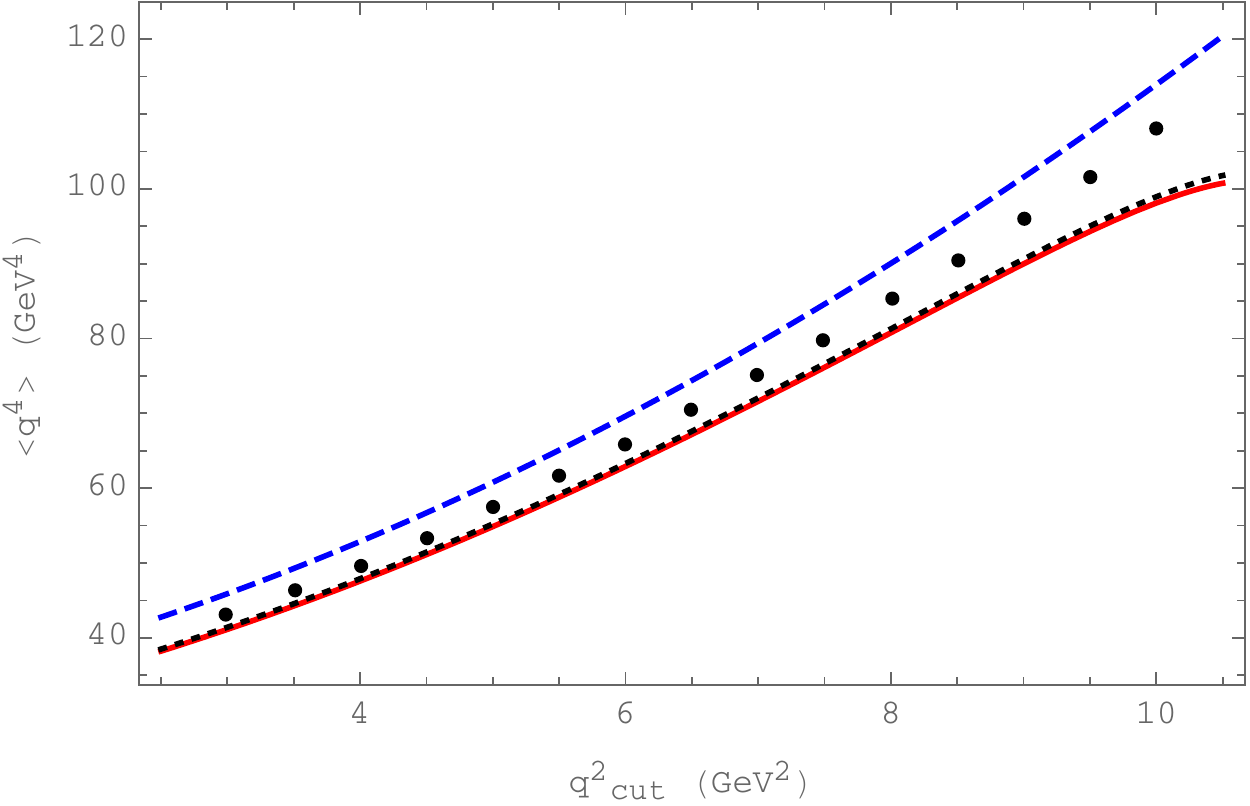}
\label{fig:subfigure3}}
\subfigure[Third moment.]{%
\includegraphics[scale=0.65]{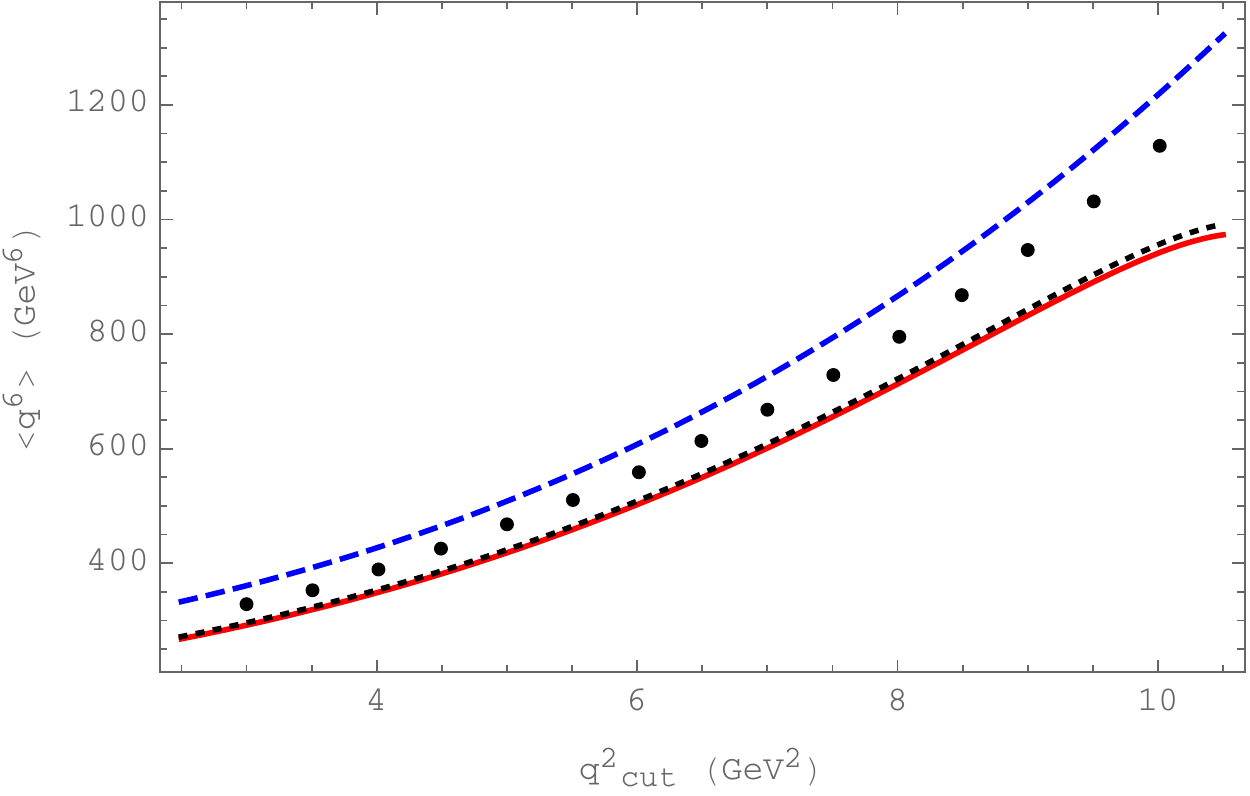}
\label{fig:subfigure2}}
\quad
\subfigure[Fourth moment.]{%
\includegraphics[scale=0.65]{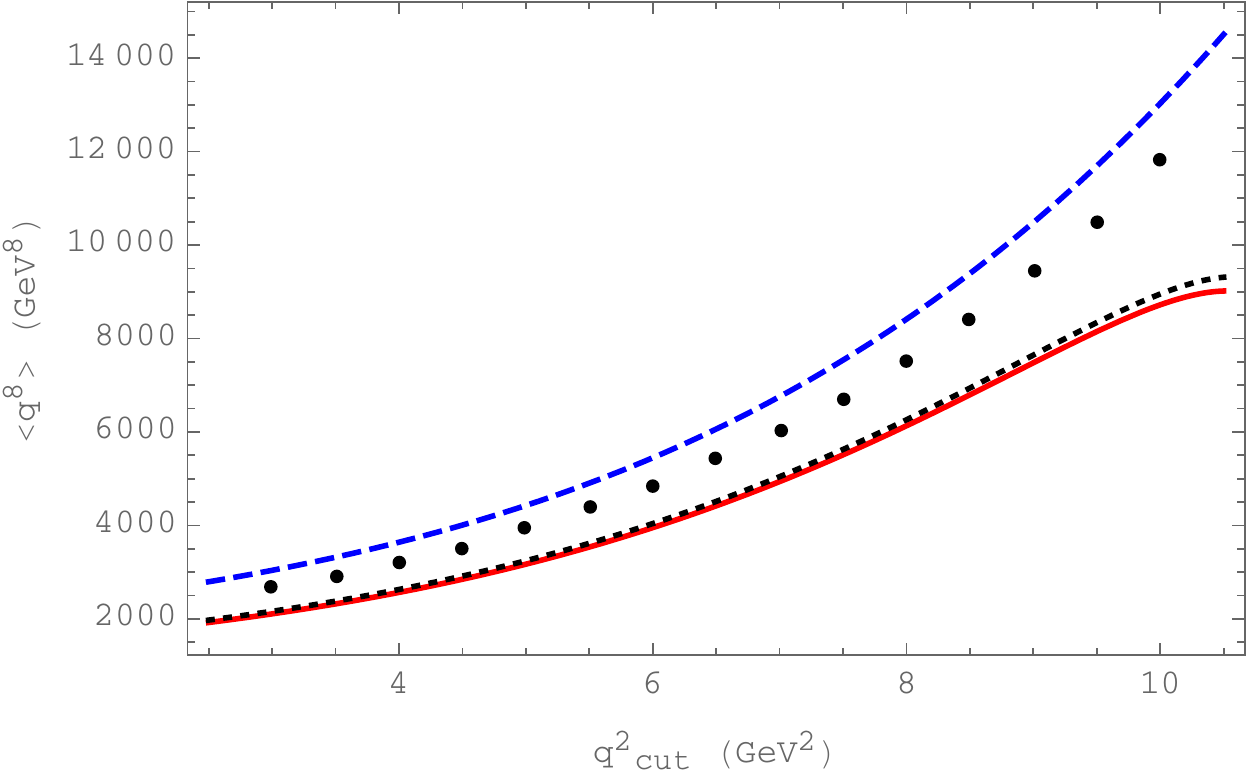}
\label{fig:subfigure4}}
\caption{$q^2$ moments normalized to the partial rate as a function of the low cut $q^2_{\rm cut}$ in the lepton pair invariant mass squared in the 
range $2.5\mbox{ GeV}^2<q^2_{\rm cut}<10.5$ GeV$^2$. 
The blue dashed line includes NLO corrections up to $1/m_b^2$, the black dotted line includes also $1/m_b^3$ corrections at LO, and 
the continuous red line includes NLO corrections up to $1/m_b^3$. Black dots correspond to experimental central values.}
\label{fig:q2mom}
\end{figure}

\begin{figure}[ht]
\centering
\subfigure[First moment.]{\includegraphics[scale=0.65]{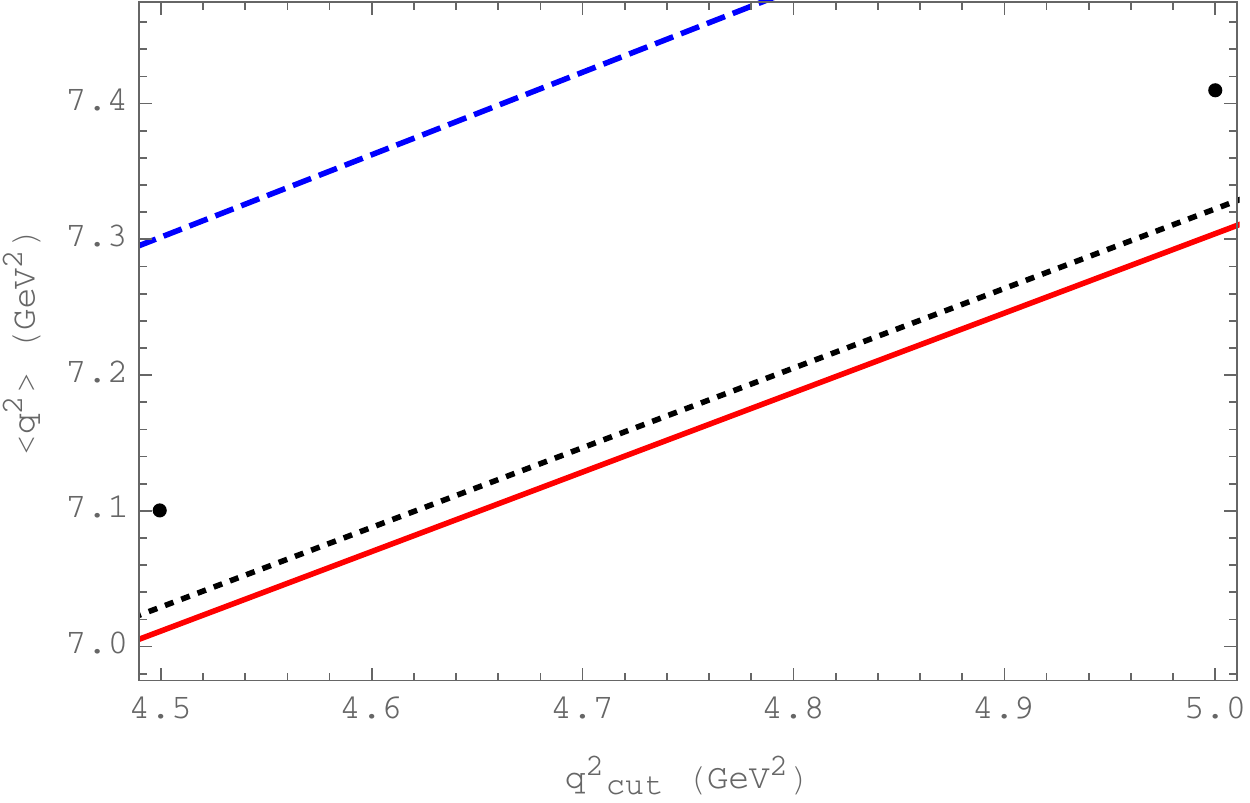}
\label{fig:subfigure1}}
\quad
\subfigure[Second moment.]{%
\includegraphics[scale=0.65]{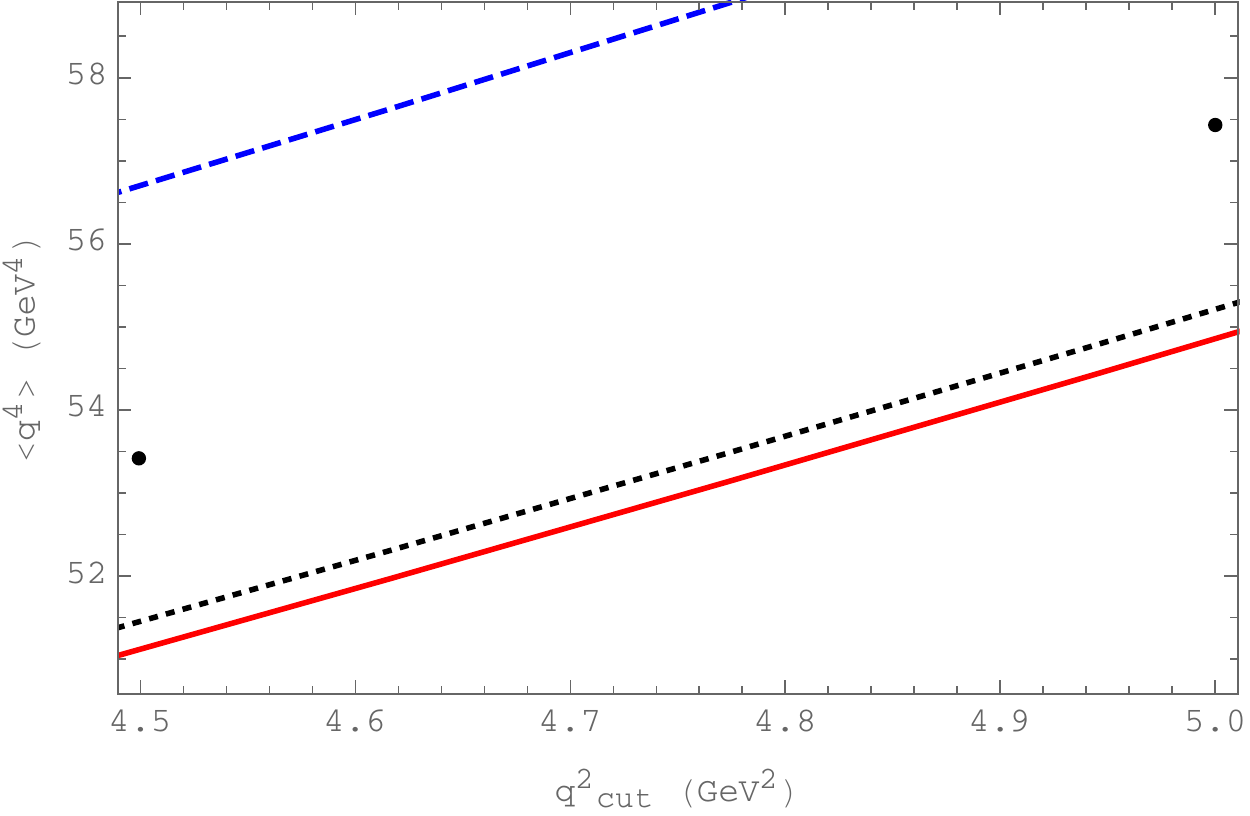}
\label{fig:subfigure3}}
\subfigure[Third moment.]{%
\includegraphics[scale=0.65]{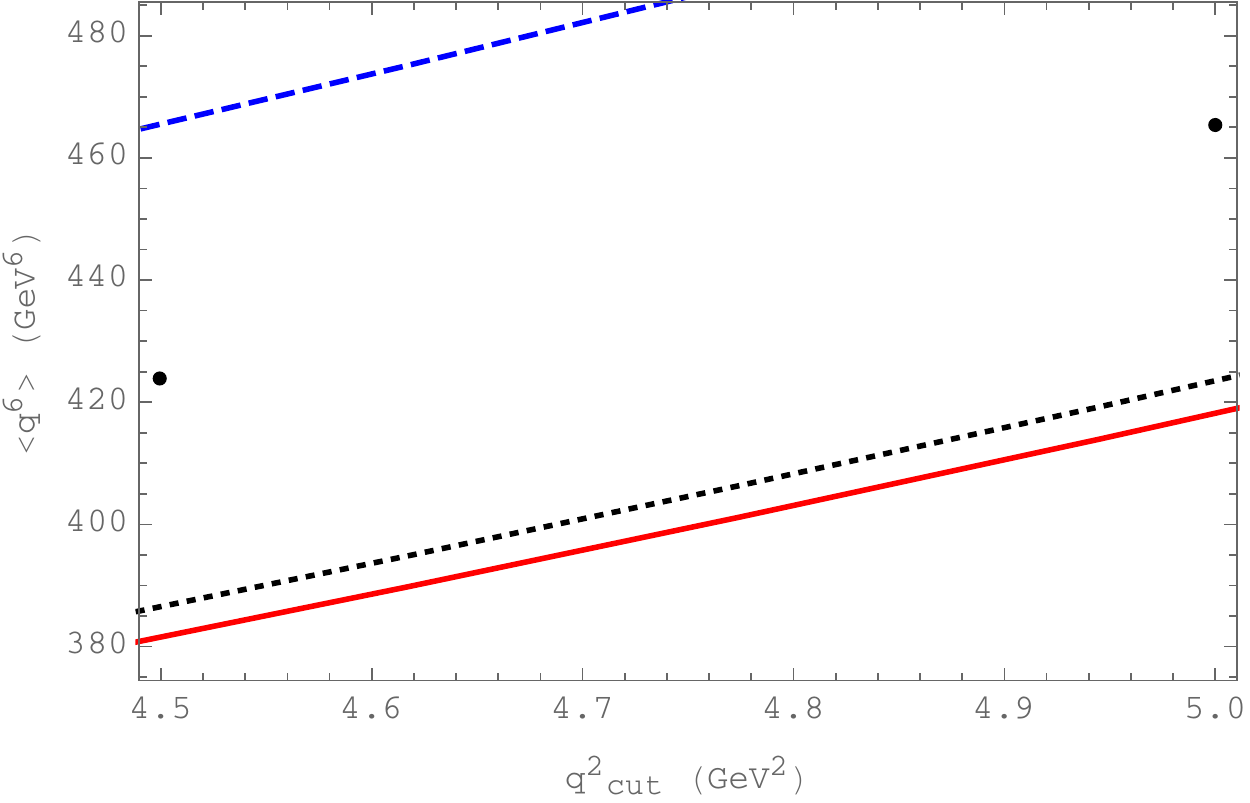}
\label{fig:subfigure2}}
\quad
\subfigure[Fourth moment.]{%
\includegraphics[scale=0.65]{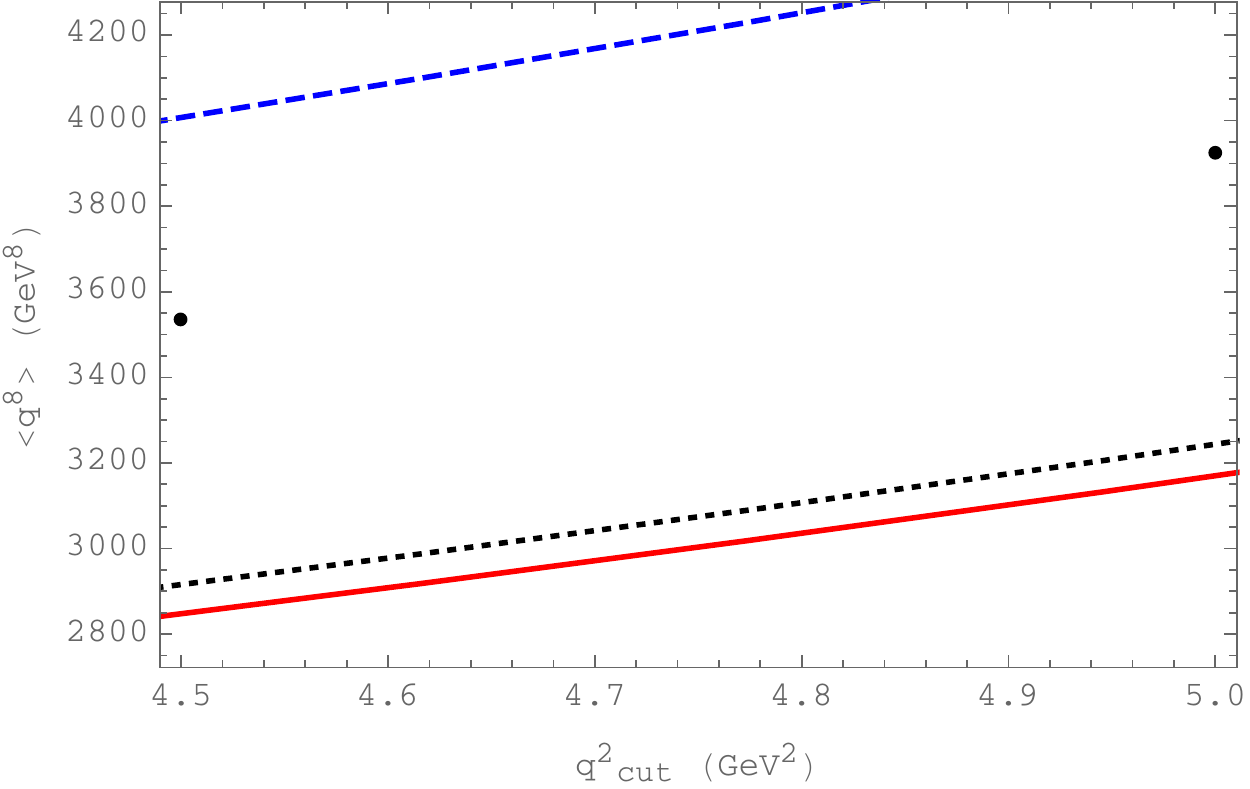}
\label{fig:subfigure4}}
\caption{$q^2$ moments normalized to the partial rate as a function of $q^2_{\rm cut}$ in 
the range $4.5\mbox{ GeV}<q^2_{\rm cut}<5\mbox{ GeV}$. 
The blue dashed line includes NLO corrections up to $1/m_b^2$, the black dotted line includes also $1/m_b^3$ corrections at LO, and 
the continuous red line includes NLO corrections up to $1/m_b^3$. Black dots correspond to experimental central values.}
\label{fig:q2momZOOM}
\end{figure}

We provide the results for the moments in the form
\begin{eqnarray}
  M_n (r_{\mbox{\scriptsize cut}}) &=& 
 \Gamma^0 |V_{cb}|^2 
 \bigg[   M_{n,0}\bigg(1 - \frac{\mu_\pi^2}{2m_b^2}\bigg)
 + M_{n,\mu_G} \bigg(\frac{\mu_G^2}{2m_b^2} - \frac{\rho_{LS}^3}{2m_b^3}\bigg)
 - M_{n,\rho_D} \frac{\rho_D^3}{2m_b^3}
 \bigg]\,,
 \label{momdef}
\end{eqnarray}
where 

\begin{eqnarray}   
 M_{n,i} &=& M_{n,i}^{\mbox{\scriptsize LO}} 
 + \frac{\alpha_s}{\pi} M_{n,i}^{\mbox{\scriptsize NLO}}
 \nonumber
 \\
 &=& M_{n,i}^{\mbox{\scriptsize LO}} 
 + \frac{\alpha_s}{\pi}\bigg(C_F M_{n,i}^{\mbox{\scriptsize NLO, F}}
 + C_A M_{n,i}^{\mbox{\scriptsize NLO, A}}\bigg)\,.
 \label{momdefLONLO}
\end{eqnarray}
To discuss the impact of our result, we show the contribution to the partial rate, 
including a lower cut in $q^2$ and its contribution to the first few (normalized) $q^2$ moments. 
Figs.~[\ref{fig:M0}], [\ref{fig:q2mom}] and [\ref{fig:q2momZOOM}] show the dependence of these quantities on the lower cut off $q^2_{\rm cut}$. 
In Figs. [\ref{fig:M0b}] and [\ref{fig:q2momZOOM}] we show the partial rate and normalized moments in a cut out range to perceive the difference between LO and NLO $1/m_b^3$ 
corrections. The NLO $1/m_b^3$ corrections represent around $1\%$ of the total contribution to the normalized moments.
Again, we use for illustration the numerical values given in table~\ref{tab:par}. 

Some specific values for the coefficients of the (not yet normalized) moments at different $q^2_{\rm cut}$ are given 
in table~\ref{tab:moms}. We observe that the NLO $1/m_b^3$ corrections represent around a $5\%$ correction of the LO $1/m_b^3$ correction. 
We also see that for larger moments there is less sensitivity to $q^2_{\rm cut}$. 
\begin{table}
\begin{center}
\begin{tabular}{ |c |c |c |c |c |c |c |c|}
 \hline
 &$q^2_{\mbox{\scriptsize cut}}$ (GeV$^2$)& $M_{n,0}^{{\rm LO}}$ & $M_{n,0}^{{\rm NLO}}$ & $M_{n,\mu_G}^{{\rm LO}}$ & $M_{n,\mu_G}^{{\rm NLO}}$ & $M_{n,\rho_D}^{{\rm LO}}$ & $M_{n,\rho_D}^{{\rm NLO}}$ \\ 
 \hline
                    &$3.0$ & $0.3835$ & $-0.6332$ & $-2.441$ & $2.18$ & $17.49$ & $14.8$ \\ 
 $M_0(q^2_{\rm cut})$ &$4.5$ & $0.2901$ & $-0.4635$ & $-2.439$ & $2.07$ & $16.89$ & $15.4$ \\ 
                    &$6.0$ & $0.2056$ & $-0.3146$ & $-2.373$ & $1.88$ & $16.19$ & $15.8$ \\
 \hline
                    &$3.0$ & $0.10938$ & $-0.1736$ & $-1.114$ & $0.823$ & $8.751$ & $8.11$ \\ 
 $M_1(q^2_{\rm cut})$ &$4.5$ & $0.09422$ & $-0.1461$ & $-1.113$ & $0.804$ & $8.651$ & $8.21$ \\ 
                    &$6.0$ & $0.07501$ & $-0.1123$ & $-1.098$ & $0.761$ & $8.492$ & $8.30$ \\
 \hline
                    &$3.0$ & $0.03527$ & $-0.05371$ & $-0.5230$ & $0.329$ & $4.584$ & $4.37$ \\ 
 $M_2(q^2_{\rm cut})$ &$4.5$ & $0.03278$ & $-0.04919$ & $-0.5228$ & $0.326$ & $4.568$ & $4.39$ \\ 
                    &$6.0$ & $0.02838$ & $-0.04146$ & $-0.5192$ & $0.316$ & $4.531$ & $4.41$ \\
 \hline 
                    &$3.0$ & $0.01255$ & $-0.01838$ & $-0.2509$ & $0.137$ & $2.454$ & $2.35$ \\ 
 $M_3(q^2_{\rm cut})$ &$4.5$ & $0.01214$ & $-0.01762$ & $-0.2509$ & $0.137$  & $2.451$ & $2.35$\\ 
                    &$6.0$ & $0.01112$  & $-0.01585$ & $-0.2500$ & $0.134$ & $2.442$ & $2.36$ \\
 \hline
                    &$3.0$ & $0.004809$ & $-0.006795$ & $-0.1223$ & $0.0588$ & $1.327$ & $1.26$ \\ 
 $M_4(q^2_{\rm cut})$ &$4.5$ & $0.004739$ & $-0.006669$ & $-0.1223$ & $0.0587$ & $1.326$ & $1.26$ \\ 
                    &$6.0$ & $0.004504$  & $-0.006257$ & $-0.1221$ & $0.0581$  & $1.324$ & $1.27$\\
 \hline
\end{tabular}
\end{center}
\caption{Numerical values for the coefficients of moments at different $q^2_{\mbox{\scriptsize cut}}$. \label{tab:moms}}
\end{table}

Some specific values for the normalized moments are given in table~\ref{tab:normmoms}

\begin{table}
\begin{center}
\begin{tabular}{ |c| c| c| c| c|}
 \hline
 $q^2_{\mbox{\scriptsize cut}}$ (GeV$^2$) & $\langle q^2\rangle$ (GeV$^2$) & $\langle q^4\rangle$ (GeV$^4$) & $\langle q^6\rangle$ (GeV$^6$) & $\langle q^8\rangle$ (GeV$^8$)\\ 
 \hline
 $3.0$ & 6.134 & 41.13 & 291.8 & 2107 \\ 
 $4.5$ & 7.011 & 51.12 & 381.5 & 2847\\ 
 $6.0$ & 7.887 & 62.90 & 502.3 & 3951\\
 \hline
\end{tabular}
\end{center}
\caption{Numerical values for normalized moments  $\langle q^{2n}\rangle = m_b^{2n}M_n/M_0$ 
at different $q^2_{\mbox{\scriptsize cut}}$. \label{tab:normmoms}}
\end{table}

\clearpage
\newpage

\section*{Conclusions}
In the present paper, we have computed the NLO corrections
up to order $1/m_b^3$ for the  
differential rate in the lepton pair invariant mass squared. Thus the current knowledge of the HQE for the 
$B \to X_c \ell \bar{\nu}$ decay includes 
the NNLO corrections to the leading term for decay
distributions
and the N$^3$LO contributions 
to the total rate, together with the NLO corrections for decay distributions
to order $1/m_b^2$ and $1/m_b^3$, while the terms at order $1/m_b^4$ and $1/m_b^5$ are known only at tree level.  

The techniques developed in~\cite{Mannel:2021ubk} can be extended to even higher orders, so the 
computation of the NLO contributions to the terms of order $1/m_b^4$ is technically possible. However, 
parametrically the largest unknown contributions are the NNLO corrections to the terms at order $1/m_b^2$ 
which are partially known for reparametrization invariant quantities due to the relation between the 
leading term and the coefficient of $\mu_\pi^2$.       

Although the corrections we have computed here are not untypically large, they will have 
a visible impact on the determination of $|V_{cb}|$. This is 
mainly due to the fact that the coefficient in front of $\rho_D$ in 
the total rate is quite large. While a detailed 
analysis will require to repeat the combined fit as e.g.
in ref.~\cite{Gambino:2013rza}, we may obtain a tendency
from an approximate formula given in eq.~(12) in this paper.

\subsection*{Acknowledgments}
This research was supported by the Deutsche Forschungsgemeinschaft 
(DFG, German Research Foundation) under grant  396021762 - TRR 257 
``Particle Physics Phenomenology after the Higgs Discovery''. 

\appendix

\section{Coefficients of the differential decay width}
\label{App:coefdifW}
In this appendix we present the coefficients of the $q^2$ spectrum relevant for phenomenology. 
We also provide these expressions in the text file "Coef.m" in Mathematica format. 

Following ref.~\cite{Mannel:2021ubk} we define  
\begin{eqnarray}
	\label{xminus}
	x_{-} &=& \frac{1}{2}\bigg(1-r+\rho - \sqrt{(1-(\sqrt{r}-\sqrt{\rho})^2)(1-(\sqrt{r}+\sqrt{\rho})^2)}\bigg)\,,
	\\
	x_{+} &=& \frac{1}{2}\bigg(1-r+\rho + \sqrt{(1-(\sqrt{r}-\sqrt{\rho})^2)(1-(\sqrt{r}+\sqrt{\rho})^2)}\bigg)\,,
	\label{xplus}
\end{eqnarray}
with $\rho=m_c^2/m_b^2$ and $r = q^2/m_b^2$. Furthermore, it is convenient to express some of the functions in terms 
of $L(x)$, which is the Roger's dilogarithm
\begin{equation}
	L(x) = \Li_2(x) + \frac{1}{2}\ln(x)\ln(1-x) = \frac{1}{2}\bigg(\frac{\pi^2}{6} + \Li_2(x) - \Li_2(1-x) \bigg)
	,\;\;\;\; 0<x<1\,.
\end{equation}
We define the full coefficients by  
\begin{eqnarray}   
 \mathcal{C}_i &=& \mathcal{C}_i^{\mbox{\scriptsize LO}} 
 + \frac{\alpha_s}{\pi}\bigg(C_F \mathcal{C}_i^{\mbox{\scriptsize NLO, F}}
 + C_A \mathcal{C}_i^{\mbox{\scriptsize NLO, A}}\bigg)\,,
\end{eqnarray}
with $i = 0,v, \mu_G, \rho_D$ and list below the coefficients of the right-hand side. The LO coefficients read

\begin{eqnarray}
 \mathcal{C}_0^{\mbox{\scriptsize LO}} &=& 48 \pi ^2 (x_{+}-x_{-})
 \Big( x_{-} x_{+} 
(3 x_{-}+3 x_{+}-8) + x_{-}(3-2 x_{-}) + x_{+} (3-2 x_{+}) \Big)\,,
 \\
 \bar{\mathcal{C}}_v^{\mbox{\scriptsize LO}} &=&  -\frac{144 \pi ^2}{x_{-}-x_{+}} 
  \bigg( 
    x_{-}^2 (2 x_{-}^2-5 x_{-}+4) 
   + x_{+}^2 (2 x_{+}^2-5 x_{+}+4)
   - x_{-}^2 x_{+}^2 (x_{-}+x_{+}-4)
   \nonumber
   \\
   &&
   - x_{-} x_{+} (3x_{-}^3 - 8x_{-}^2 + 7x_{-} + 3 x_{+}^3  - 8 x_{+}^2  + 7 x_{+}) 
   \bigg)\,,
 \\
 \mathcal{C}_{\mu_G}^{\mbox{\scriptsize LO}} &=& 
  \frac{48\pi^2}{x_{-}-x_{+}} \bigg(
   x_{-}(10x_{-}^3 - 15x_{-}^2 + 4) 
  + x_{+}(10x_{+}^3 - 15x_{+}^2 + 4)
  + 3x_{-}^2 x_{+}^2 (x_{-}+x_{+}-4)
  \nonumber
  \\
  &&
  - x_{-} x_{+}(15 x_{-}^3-28 x_{-}^2+9 x_{-}+15 x_{+}^3-28 x_{+}^2+9 x_{+})
   \bigg)\,,
   \\
   \mathcal{C}_{\rho_D}^{\mbox{\scriptsize LO}} &=& 
 \frac{16 \pi ^2 }{(x_{-}-x_{+})^3}
 \bigg( 
  x_{-}^2 (-10 x_{-}^4+63x_{-}^3-120 x_{-}^2+104 x_{-}-48)
 \nonumber
 \\
 &&
 + x_{+}^2 (-10x_{+}^4+63 x_{+}^3-120 x_{+}^2+104 x_{+}-48)
 -22 x_{-}^3 x_{+}^3 (3 x_{-}+3 x_{+}-16)
 \nonumber
 \\
 &&
 + x_{-}^2 x_{+}^2 (3 x_{-}^3+154x_{-}^2-570 x_{-}+3 x_{+}^3+154 x_{+}^2-570 x_{+}+720)
 \nonumber
 \\
 &&
  + x_{-} x_{+} (15 x_{-}^5-80 x_{-}^4+27 x_{-}^3+240x_{-}^2-344 x_{-}+15 x_{+}^5-80 x_{+}^4+27 x_{+}^3
   \nonumber
 \\
 &&
  +240 x_{+}^2-344 x_{+}+192)
  \bigg)
  \,.
\end{eqnarray}
The NLO coefficients read

\begin{eqnarray}
\mathcal{C}_0^{\mbox{\scriptsize NLO, A}} &=& 0 \,,
\\
 \mathcal{C}_0^{\mbox{\scriptsize NLO, F}} &=& 
   - 24\pi^2 \bigg\{
    \frac{1}{2}(x_{-}-x_{+}) \Big( x_{-} x_{+}(8 x_{-} x_{+} + 3 x_{-}  + 3 x_{+} - 28) + 3x_{-}(1 - 2x_{-}) 
   \nonumber
   \\
   &&   
   + 3x_{+}(1 - 2x_{+}) + 8 \Big)
   + 4\Big(  x_{-} x_{+}( 3x_{-} + 3x_{+} - 8) + x_{-}(3 - 2x_{-}) + x_{+}(3 - 2x_{+}) \Big)
   \nonumber
   \\
   &&   
   \times\bigg[  
    (x_{-}+x_{+})\bigg( 2L\bigg(1-\frac{x_{-}}{x_{+}}\bigg) + L(x_{-}) - L(x_{+}) \bigg)
  - 2(x_{-}-x_{+})\ln (x_{+}-x_{-}) \bigg]
   \nonumber
   \\
   &&
   + x_{-} \bigg(  x_{-}^2 x_{+}^2(4x_{+} - 14) + x_{-}x_{+}( -14x_{+}^2 + 34x_{+} + 26x_{-} - 40)
   \nonumber
   \\
   &&
                + x_{-}(15 - 12x_{-}) + x_{+}(5x_{+}^2 - 4x_{+} + 6) \bigg)\ln(x_{-})
   \nonumber
   \\
   &&
   - x_{+}\bigg( x_{-}^2 x_{+}^2(4x_{-} - 14) + x_{-}x_{+}( -14x_{-}^2 + 34x_{-} + 26x_{+} - 40)
   \nonumber
   \\
   &&
   + x_{+}(15 - 12x_{+}) + x_{-}(5 x_{-}^2 - 4x_{-} + 6) \bigg) \ln(x_{+}) 
   \nonumber
   \\
   &&
   - \bigg( 
    4x_{-}^3 x_{+}^3 
   - x_{-}^2 x_{+}^2(14x_{-} + 14x_{+} - 28)
   + x_{-} x_{+}( 14x_{+}^2 - 12x_{+} + 2x_{-}^2 + 12x_{-} - 28 )
   \nonumber
   \\
   &&
   + x_{-}( 4x_{-}^2 - 14x_{-} + 14) 
   + x_{+}(-4x_{+}^2 - 2x_{+} + 14) - 4 \bigg)\ln (1-x_{-})
   \nonumber
   \\
   &&
   +\bigg( 
   4x_{-}^3 x_{+}^3 
   - x_{-}^2 x_{+}^2 (14 x_{-} + 14x_{+} - 28)
   + x_{-}x_{+}(14 x_{-}^2 - 12x_{-} + 2x_{+}^2  + 12x_{+}  - 28)
   \nonumber
   \\
   &&
   + x_{+}( 4x_{+}^2 - 14x_{+} + 14 )
   + x_{-}(-4x_{-}^2 - 2x_{-} + 14) - 4\bigg) \ln(1-x_{+})
   \bigg\}
   \,,
\end{eqnarray}

\begin{eqnarray}
 \bar{\mathcal{C}}_v^{\mbox{\scriptsize NLO, A}} &=& 0
 \\
 \bar{\mathcal{C}}_v^{\mbox{\scriptsize NLO, F}} &=& 
   -24\pi^2 \bigg\{
   -4\Big( 18 x_{-}^2 x_{+}^2 + x_{-} x_{+} (9 x_{-}^2-32 x_{-}+9 x_{+}^2-32
   x_{+}+28) -6 x_{-} (x_{-}-1)^2 
   \nonumber
   \\
   &&
   -6 x_{+} (x_{+}-1)^2 \Big)
   \bigg(  2L\left(1-\frac{x_{-}}{x_{+}}\right) + L(x_{-}) - L(x_{+})\bigg)
   \nonumber
   \\
   &&
   +4 (x_{-}-x_{+}) (5 x_{-}+5 x_{+}-6) (3 x_{-} x_{+}-2 x_{-}-2 x_{+}+1) \ln (x_{+}-x_{-})
   \nonumber
   \\
   &&
   + \frac{1}{2(x_{-}-x_{+})} \bigg( 
   48 x_{-}^3 x_{+}^3
   + x_{-}^2 x_{+}^2 ( -24x_{-}^2 + 87x_{-} -24x_{+}^2+87 x_{+}-352) 
   \nonumber
   \\
   &&
   + 2x_{-}^2 (3 x_{-}^2+5 x_{-}-17 ) 
   + x_{-} x_{+} ( 9x_{-}^3 - 118 x_{-}^2 + 278 x_{-} + 9x_{+}^3 -118 x_{+}^2 + 278x_{+} -124)
   \nonumber
   \\
   &&
   +2 x_{+}^2 (3x_{+}^2+5 x_{+}-17)
   \bigg)
   \nonumber
   \\
   &&
   + (x_{-}-1) \bigg( 6 x_{-}^2 x_{+}^2 (2x_{+}-7) + x_{-} x_{+} (12 x_{-}-30 x_{+}^2+59 x_{+}-4) + 
   x_{-} (8x_{-}-13)
   \nonumber
   \\
   &&
   + x_{+}(12 x_{+}^2-13 x_{+}-6) +5 \bigg) \ln (1-x_{-})
   \nonumber
   \\
   &&
   - (x_{+}-1) \bigg( 6x_{-}^2 x_{+}^2 (2 x_{-}-7) + x_{-} x_{+} (12 x_{+}-30 x_{-}^2+59 x_{-}-4) 
   + x_{+} (8 x_{+}-13) 
   \nonumber
   \\
   &&
   + x_{-}(12 x_{-}^2-13 x_{-}-6) +5\bigg)\ln (1-x_{+}) 
   \nonumber
   \\
   &&
   - \frac{x_{-}}{(x_{-}-x_{+})^2} \bigg( 
    x_{-}^2 (-32 x_{-}^2+29 x_{-}+12)
   +3 x_{+}^2 (5 x_{+}^3-2 x_{+}^2-14 x_{+}+8) 
   \nonumber
   \\
   &&
   + x_{-}^3 x_{+}^3 ( 12 x_{-} - 24 x_{+} + 42)
   + x_{-}^2 x_{+}^2 \left( - 43x_{-} - 42x_{-}^2 - 175x_{+} + 42x_{+}^2 + 12x_{+}^3 + 266 \right)
   \nonumber
   \\
   && 
   + x_{-} x_{+} (- 136 x_{-} - 34 x_{-}^2 + 72 x_{-}^3 -139x_{+} + 94x_{+}^2 + 35x_{+}^3 - 42x_{+}^4 + 60)
   \bigg) \ln (x_{-})
   \nonumber
   \\
   &&
   + \frac{x_{+}}{(x_{-}-x_{+})^2}\bigg(
   x_{+}^2 (-32 x_{+}^2 + 29x_{+}+12)
   + 3x_{-}^2 (5x_{-}^3-2 x_{-}^2-14 x_{-}+8) 
   \nonumber
   \\
   &&   
   + x_{-}^3 x_{+}^3 (12 x_{+} - 24x_{-} + 42)
   + x_{-}^2 x_{+}^2 (-43 x_{+} - 42 x_{+}^2 -175x_{-} + 42x_{-}^2 + 12x_{-}^3 + 266)
   \nonumber
   \\
   &&
   +x_{-} x_{+} (-136 x_{+} -34 x_{+}^2 +72x_{+}^3 -139x_{-} + 94x_{-}^2 + 35 x_{-}^3 - 42 x_{-}^4 + 60)   
   \bigg) \ln (x_{+})
   \bigg\}
\end{eqnarray}

\begin{eqnarray}
 \mathcal{C}_{\mu_G}^{\mbox{\scriptsize NLO, A}} &=& 
  - 16\pi^2 \bigg\{ 
      - \frac{1}{4(x_{-}-x_{+})} \bigg( 
   9x_{-}^2 x_{+}^2 (x_{-}+x_{+}+12)
   \nonumber
   \\
   &&
   + x_{-} x_{+} (135 x_{-}^3-348 x_{-}^2+105 x_{-}+135x_{+}^3-348 x_{+}^2+105 x_{+}+40)
   \nonumber
   \\
   &&
   - x_{-} (90 x_{-}^3-183 x_{-}^2 + 20x_{-}+48)
   - x_{+} (90 x_{+}^3-183 x_{+}^2+20 x_{+}+48) 
   \bigg)
   \nonumber
   \\
   &&
   - 2\bigg( 9x_{-}^2 x_{+}^2 - 2x_{-}x_{+} (7 x_{-}+7 x_{+}-8) + 2x_{-} (4 x_{-}-3) + 2x_{+}(4x_{+}-3) - 1 \bigg) 
   \nonumber
   \\
   &&
   \times \bigg( 2L\bigg(1-\frac{x_{-}}{x_{+}}\bigg) + L(x_{-}) - L(x_{+}) \bigg)
   \nonumber
   \\
   &&   
   + \frac{x_{-}-x_{+}}{x_{-} x_{+}} \bigg( x_{-}^2 x_{+}^2 (9x_{-}+9 x_{+}-20) 
   - 2x_{-} x_{+} (3 x_{-}^2 - x_{-} + 3x_{+}^2 - x_{+} - 6) 
   \nonumber
   \\
   &&
   + x_{-} (6x_{-}-7) + x_{+} (6x_{+}-7) \bigg) \ln (x_{+}-x_{-})
   \nonumber
   \\
   &&
   -\frac{x_{-}-1}{2x_{-}}\bigg( 17 x_{-}^2 x_{+}^2 + x_{-} x_{+} (9 x_{-}^2-39 x_{-}-17x_{+}+29)
   \nonumber
   \\
   &&
   - x_{-} (6x_{-}^2 - 19x_{-} + 11) 
   + x_{+} (6 x_{+}-7) \bigg) \ln(1-x_{-})
   \nonumber
   \\
   &&
   +\frac{x_{+}-1}{2x_{+}} \bigg( 17 x_{-}^2 x_{+}^2 + x_{-} x_{+} (9x_{+}^2-39 x_{+}-17 x_{-}+29)
   \nonumber
   \\
   &&
   - x_{+} (6x_{+}^2 - 19x_{+} + 11) + x_{-} (6 x_{-}-7)  \bigg)\ln (1-x_{+})
   \nonumber
   \\
   &&
   +\frac{x_{-}}{2x_{+} (x_{-}-1) (x_{-}-x_{+})^2} \bigg(
   x_{-}^3 x_{+}^3 (62 x_{-}+47 x_{+}-266) 
   \nonumber
   \\
   &&
   - x_{-}^2 x_{+}^2 (9 x_{-}^3 + 17 x_{-}^2 - 142 x_{-} - 44 x_{+}^3 + 173 x_{+}^2 - 311 x_{+} + 96)
   \nonumber
   \\
   &&
   + x_{-} x_{+} (6 x_{-}^4-9 x_{-}^3-13 x_{-}^2+6 x_{-}-78
   x_{+}^4+188 x_{+}^3-81 x_{+}^2-90 x_{+}+16)
   \nonumber
   \\
   &&
   - 2 x_{-}^2 (6x_{-}^2-13 x_{-}+7) 
   + 2x_{+}^2 (20 x_{+}^3-43 x_{+}^2+5 x_{+}+23) \bigg)
   \ln (x_{-})
   \nonumber
   \\
   &&
   -\frac{x_{+}}{2x_{-} (x_{+}-1) (x_{-}-x_{+})^2} \bigg( 
   x_{-}^3 x_{+}^3 (62 x_{+}+47 x_{-}-266)
   \nonumber
   \\
   &&
   - x_{-}^2 x_{+}^2 (9x_{+}^3 + 17 x_{+}^2 - 142 x_{+} -44 x_{-}^3 + 173 x_{-}^2-311 x_{-}+96)
   \nonumber
   \\
   &&
   + x_{-} x_{+} (6 x_{+}^4 - 9x_{+}^3 - 13 x_{+}^2 + 6 x_{+} - 78 x_{-}^4 + 188 x_{-}^3 - 81 x_{-}^2-90 x_{-}+16)
   \nonumber
   \\
   &&
   - 2x_{+}^2 (6 x_{+}^2-13 x_{+}+7) 
   + 2x_{-}^2 (20 x_{-}^3-43 x_{-}^2+5x_{-}+23) \bigg) \ln (x_{+})
   \bigg\}
   \,,
\end{eqnarray}

\begin{eqnarray}
 \mathcal{C}_{\mu_G}^{\mbox{\scriptsize NLO, F}} &=& 
   -16 \pi^2\bigg\{ 
    \frac{1}{4} (x_{-}-x_{+}) \bigg( x_{-}x_{+} ( 120x_{-} x_{+} + 117x_{-} + 117 x_{+}-332) 
   \nonumber
   \\
   && 
   + (85-138 x_{-}) x_{-}
   + (85-138 x_{+}) x_{+}+48 \bigg)
   \nonumber
   \\
   && 
   + 2\bigg(66 x_{-}^2 x_{+}^2 + x_{-} x_{+} (45 x_{-}^2-122 x_{-}+45 x_{+}^2-122x_{+}+86) 
   - x_{-} (30x_{-}^2 - 37 x_{-} + 4)
   \nonumber
   \\
   &&
   - x_{+} (30 x_{+}^2 - 37x_{+} + 4) - 4\bigg)
   \bigg( 2L\bigg(1-\frac{x_{-}}{x_{+}}\bigg) + L(x_{-}) - L(x_{+}) \bigg)
   \nonumber
   \\
   &&
   - \frac{8(x_{-}-x_{+})}{x_{-} x_{+}} \bigg( x_{-}^2 x_{+}^2 (15x_{-}+15 x_{+}-32) 
   - x_{-} x_{+} (10x_{-}^2 - 11 x_{-} + 10 x_{+}^2 - 11x_{+} )
   \nonumber
   \\
   &&
   - (x_{-}-1) x_{-} - (x_{+}-1) x_{+}\bigg) \ln(x_{+}-x_{-})
   \nonumber
   \\
   &&
   + \frac{x_{+}-1}{x_{+}}\bigg( 4(x_{-}-1) x_{-} + x_{+}(10 x_{+}^2-17 x_{+} + 8)
   + 30 x_{-}^3 x_{+}^3 
   \nonumber
   \\
   &&
   - x_{-}^2 x_{+}^2 (75x_{-} + 105x_{+} -133) 
   + x_{-} x_{+} (30 x_{-}^2-13 x_{-}+45 x_{+}^2-13 x_{+}-33)
    \bigg) \ln(1-x_{+})
   \nonumber
   \\
   &&
   - \frac{x_{-}-1}{x_{-}} \bigg( 4(x_{+}-1) x_{+} + x_{-} (10x_{-}^2-17 x_{-}+8) + 30 x_{-}^3 x_{+}^3  
   \nonumber
   \\
   &&
   - x_{-}^2 x_{+}^2 (75x_{+} + 105 x_{-} - 133)
   + x_{-} x_{+} ( 30x_{+}^2 -13 x_{+} + 45x_{-}^2 - 13x_{-} -33) \bigg) \ln (1-x_{-})
   \nonumber
   \\
   &&
   + \frac{x_{-}}{2 x_{+} (x_{-}-1) (x_{-}-x_{+})^2} \bigg( 30 x_{-}^4 x_{+}^4 (2 x_{-}-4 x_{+}+5)-16 (x_{-}-1)^2 x_{-}^2 
   \nonumber
   \\
   &&
   - x_{-}^3 x_{+}^3 (210x_{-}^2 + 136x_{-} - 60x_{+}^3 - 330x_{+}^2 + 901x_{+} - 1486)
   \nonumber
   \\
   &&
   + x_{-}^2 x_{+}^2 (330 x_{-}^3-326 x_{-}^2-464 x_{-}-270 x_{+}^4-154 x_{+}^3+1191 x_{+}^2-1759 x_{+}+452)
   \nonumber
   \\
   &&
   -x_{-} x_{+} (140 x_{-}^4-217 x_{-}^3+x_{-}^2+84 x_{-}-285 x_{+}^5 + 24x_{+}^4+638 x_{+}^3-531 x_{+}^2-228 x_{+}+4)
   \nonumber
   \\
   &&
   - x_{+}^2( 75x_{+}^4 + 44x_{+}^3 - 186x_{+}^2 - 16x_{+} + 172) \bigg) \ln (x_{-})
   \nonumber
   \\
   &&
   - \frac{x_{+}}{2x_{-} (x_{+}-1)(x_{-}-x_{+})^2} \bigg( 30x_{-}^4 x_{+}^4 (2x_{+} - 4x_{-} + 5) - 16 (x_{+}-1)^2 x_{+}^2
   \nonumber
   \\
   &&
   - x_{-}^3 x_{+}^3 (210x_{+}^2 + 136x_{+} - 60x_{-}^3 - 330x_{-}^2 + 901x_{-} - 1486)
   \nonumber
   \\
   &&
   + x_{-}^2 x_{+}^2 ( 330x_{+}^3 - 326x_{+}^2 - 464x_{+} - 270x_{-}^4 - 154x_{-}^3 + 1191x_{-}^2 - 1759x_{-} + 452)
   \nonumber
   \\
   && 
   - x_{-} x_{+} (140x_{+}^4 - 217x_{+}^3 + x_{+}^2 + 84x_{+} -285x_{-}^5 + 24x_{-}^4 + 638x_{-}^3 - 531x_{-}^2 - 228x_{-} + 4)
   \nonumber
   \\
   && 
   - x_{-}^2 (75 x_{-}^4+44 x_{-}^3-186 x_{-}^2-16x_{-}+172)\bigg) \ln(x_{+})
   \bigg\}
   \,,
\end{eqnarray}

\begin{eqnarray}
\mathcal{C}_{\rho_D}^{\mbox{\scriptsize NLO, A}} &=& 
-16\pi^2 \bigg\{ 
 \frac{1}{9(x_{-}-1) x_{-} (x_{-}-x_{+})^3 (x_{+}-1) x_{+}}
\bigg(
   -2 x_{-}^5 x_{+}^5 (555 x_{-}+555 x_{+}-3794)  
   \nonumber
   \\
   &&
   + 51 x_{+}^4 (2 x_{+}^3-9 x_{+}^2 + 13x_{+}-6) 
   + 2 x_{-}^4 x_{+}^4 (150 x_{-}^3
   +1493 x_{-}^2-8881 x_{-}+150 x_{+}^3
   \nonumber
   \\
   &&
   +1493x_{+}^2-8881 x_{+}+18512)  
   + x_{-}^3 x_{+}^3 (222 x_{-}^5-1862 x_{-}^4+103 x_{-}^3+14137x_{-}^2-32312 x_{-}
   \nonumber
   \\
   &&
   +222 x_{+}^5-1862 x_{+}^4+103 x_{+}^3+14137 x_{+}^2-32312 x_{+}+28792) 
   + x_{-}^2 x_{+}^2(-370 x_{-}^6+2571 x_{-}^5
   \nonumber
   \\
   &&
   -4484 x_{-}^4+62 x_{-}^3+8379 x_{-}^2-8942x_{-}-370 x_{+}^6+2571 x_{+}^5-4484 x_{+}^4+62 x_{+}^3+8379 x_{+}^2
   \nonumber
   \\
   &&
   -8942 x_{+}+2748) 
   + x_{-} x_{+}( 148 x_{-}^7 - 1105 x_{-}^6 + 2928 x_{-}^5 - 3488x_{-}^4 + 1571 x_{-}^3 - 48x_{-}^2 
   \nonumber
   \\
   &&
   + 148x_{+}^7 - 1105x_{+}^6 + 2928 x_{+}^5 - 3488x_{+}^4 + 1571x_{+}^3 -48x_{+}^2 ) 
   \nonumber
   \\
   &&   
   +51 x_{-}^4 (2 x_{-}^3-9 x_{-}^2+13 x_{-}-6)
   \bigg)
   \nonumber
   \\
   && 
   + \frac{2}{3 x_{-}^2 (x_{-}-x_{+})^3 x_{+}^2}
   \bigg(
   17(2 x_{-}-3) x_{-}^6 - 2 x_{+}^5 x_{-}^5 (33 x_{-}+33 x_{+}-206) 
   -4 x_{+}^4 x_{-}^4 (6 x_{-}^3
   \nonumber
   \\
   &&
   -43 x_{-}^2
   +162 x_{-}+6 x_{+}^3-43 x_{+}^2+162 x_{+}-159) 
   +x_{+}^3 x_{-}^3 (42x_{-}^5-110 x_{-}^4+90 x_{-}^3+279 x_{-}^2
   \nonumber
   \\
   &&
   -208 x_{-}
   +42 x_{+}^5-110 x_{+}^4+90 x_{+}^3+279x_{+}^2-208 x_{+}+168) 
   -x_{+}^2 x_{-}^2 (28 x_{-}^6-78 x_{-}^5+78 x_{-}^4
   \nonumber
   \\
   &&
   +88x_{-}^3+93 x_{-}^2
   + 28 x_{+}^6-78 x_{+}^5+78 x_{+}^4+88 x_{+}^3+93x_{+}^2 ) 
   + x_{+} x_{-} (-39 x_{-}^6+22 x_{-}^5+108 x_{-}^4 
   \nonumber
   \\
   &&
   -39 x_{+}^6 + 22x_{+}^5
   +108x_{+}^4 )
   +17 x_{+}^6 (2 x_{+}-3)
   \bigg) \ln (x_{+}-x_{-})
   \nonumber
   \\
   &&
   -\frac{2}{(x_{-}-x_{+})^3}
   \bigg( 
   x_{-}^3 x_{+}^3 (-9 x_{-}-9 x_{+}+68) 
   -2 x_{-}^2 x_{+}^2 (21 x_{-}^3-77 x_{-}^2+138 x_{-}+21 x_{+}^3-77 x_{+}^2
   \nonumber
   \\
   &&
   +138 x_{+}-180) 
   + x_{+}^2 (-18x_{+}^4+51 x_{+}^3-60 x_{+}^2+47 x_{+}-24) 
   +x_{-} x_{+} (27 x_{-}^5-50 x_{-}^4-15x_{-}^3
   \nonumber
   \\
   &&
   +120 x_{-}^2-167 x_{-}+27 x_{+}^5-50 x_{+}^4-15 x_{+}^3+120 x_{+}^2-167 x_{+}+96)
   +x_{-}^2 (-18 x_{-}^4+51 x_{-}^3
   \nonumber
   \\
   &&
   -60 x_{-}^2+47 x_{-}-24)
   \bigg) 
   \ln \left(\frac{\mu}{m_b}\right)
   \nonumber
   \\
   &&
   + 2(15 x_{-}^2 x_{+}^2 - 6 x_{-} x_{+} (3 x_{-}+3x_{+}-5)  + 2x_{+} (4 x_{+}-9) + 2 x_{-} (4 x_{-}-9)+11)
   \nonumber
   \\
   &&
   \times\bigg(
   2L\bigg(1-\frac{x_{-}}{x_{+}}\bigg) + L(x_{-}) - L(x_{+})
   \bigg)
   \nonumber
\end{eqnarray}
\begin{eqnarray}
 &&  +\frac{1}{3 (x_{-}-x_{+})^3}\bigg[
   -\frac{(x_{-}-1)}{x_{-}^2}
   \bigg(
   -63x_{+}^4 x_{-}^4 + x_{-}^4 (-28 x_{-}^3+56 x_{-}^2-46 x_{-}+15)  
   \nonumber
   \\
   &&
   + x_{+}^3 x_{-}^3 (-3x_{-}^2+203 x_{-}+33 x_{+}^2-85 x_{+}-31)  
   + x_{+}^2 x_{-}^2 (-57 x_{-}^4+137x_{-}^3-277 x_{-}^2
   \nonumber
   \\
   &&
   +41 x_{-}+11 x_{+}^3+5 x_{+}^2+41 x_{+}+78)  
   + x_{+} x_{-} ( 42x_{-}^6 - 46x_{-}^5 - 46 x_{-}^4 + 161 x_{-}^3 - 84 x_{-}^2 
   \nonumber
   \\
   &&
   + 5 x_{+}^4 + 29x_{+}^3-108x_{+}^2 )  
   +17x_{+}^4 (3-2 x_{+})  \bigg)\ln (1-x_{-})
   \nonumber
   \\
   &&
   -\frac{(x_{+}-1)}{x_{+}^2}
   \bigg( 
   -63 x_{+}^4 x_{-}^4 +17 x_{-}^4 (3-2 x_{-})  
   + x_{+}^3 x_{-}^3 (33 x_{-}^2-85 x_{-}-3 x_{+}^2
   \nonumber
   \\
   &&
   +203 x_{+}-31)  
   + x_{+}^2 x_{-}^2 (-57 x_{+}^4+137x_{+}^3-277 x_{+}^2+41 x_{+}+11 x_{-}^3+5 x_{-}^2+41 x_{-}+78 ) 
   \nonumber
   \\
   &&
   +x_{+} x_{-} ( 5x_{-}^4+29 x_{-}^3-108 x_{-}^2 
   + 42 x_{+}^6-46 x_{+}^5-46 x_{+}^4+161x_{+}^3-84x_{+}^2 )  
   \nonumber
   \\
   &&
   + x_{+}^4 (-28 x_{+}^3+56 x_{+}^2-46 x_{+}+15) 
   \bigg)\ln (1-x_{+})
   \bigg]
   \nonumber
   \\
   &&
   + \frac{1}{3 (x_{-}-x_{+})^4}
   \bigg[
   \frac{1}{x_{-}^2 (x_{+}-1)^2} \bigg( 
   -552 x_{-}^6 x_{+}^6 + 34 (x_{+}-1)^2 (2 x_{+}-3) x_{+}^5 
   + 2 x_{-}^5  x_{+}^5 (96 x_{-}^2
   \nonumber
   \\
   &&
   +839 x_{-}+75 x_{+}^2+482x_{+}-2305) 
   -x_{-}^4 x_{+}^4 (3 x_{-}^4+554 x_{-}^3+2116 x_{-}^2-7850 x_{-}+69x_{+}^4
   \nonumber
   \\
   &&
   -25 x_{+}^3+1204 x_{+}^2-4766 x_{+}+7948)  
   + x_{-}^3 x_{+}^3 (42 x_{+}^6-122x_{+}^5+420 x_{+}^4-474 x_{+}^3+106 x_{+}^2
   \nonumber
   \\
   &&
   +374 x_{+}+5 x_{-}^5+678 x_{-}^4+1088 x_{-}^3-6530x_{-}^2+7024 x_{-}-862) 
   + x_{-}^2 x_{+}^2(-22 x_{-}^6-426 x_{-}^5
   \nonumber
   \\
   &&
   +283 x_{-}^4+2116x_{-}^3-2870 x_{-}^2+528 x_{-} - 28 x_{+}^7+86 x_{+}^6-48 x_{+}^5-315 x_{+}^4+171x_{+}^3+540 x_{+}^2
   \nonumber
   \\
   &&
   -408x_{+} )  
   + x_{-} x_{+} (42 x_{-}^7+68 x_{-}^6-429 x_{-}^5 + 154x_{-}^4+246 x_{-}^3 + 318 x_{+}^3 - 660 x_{+}^4
   \nonumber
   \\
   &&
    + 288 x_{+}^5 + 132 x_{+}^6 - 78 x_{+}^7 ) 
   -6 x_{-}^5 (4 x_{-}^3-9 x_{-}^2-3 x_{-}+9)
   \bigg) \ln (x_{+})
   \nonumber
   \\
   &&
   -\frac{1}{(x_{-}-1)^2 x_{+}^2}
   \bigg(
   -552 x_{-}^6 x_{+}^6 
   + 2 x_{-}^5 x_{+}^5 (75 x_{-}^2+482 x_{-}+96 x_{+}^2+839x_{+}-2305)  
   \nonumber
   \\
   &&
   -6 x_{+}^5 (4 x_{+}^3-9 x_{+}^2-3 x_{+}+9) 
   -x_{-}^4 x_{+}^4 ( 69x_{-}^4-25 x_{-}^3+1204 x_{-}^2-4766 x_{-}+3 x_{+}^4+554 x_{+}^3
   \nonumber
   \\
   &&
   +2116 x_{+}^2-7850x_{+}+7948)  
   + x_{-}^3 x_{+}^3 (42 x_{-}^6-122 x_{-}^5+420 x_{-}^4-474 x_{-}^3+106x_{-}^2+374 x_{-}
   \nonumber
   \\
   &&
   +5 x_{+}^5+678 x_{+}^4+1088 x_{+}^3-6530 x_{+}^2+7024 x_{+}-862)
   +x_{-}^2 x_{+}^2 (-28 x_{-}^7+86 x_{-}^6-48 x_{-}^5
   \nonumber
   \\
   &&
   -315 x_{-}^4+171 x_{-}^3+540 x_{-}^2-408x_{-}-22 x_{+}^6-426 x_{+}^5+283 x_{+}^4+2116 x_{+}^3-2870 x_{+}^2
   \nonumber
   \\
   &&
   +528x_{+} ) 
   +x_{-} x_{+} ( -78 x_{-}^7+132 x_{-}^6+288 x_{-}^5-660 x_{-}^4+318 x_{-}^3 
   + 42 x_{+}^7+68 x_{+}^6 -429 x_{+}^5
   \nonumber
   \\
   && 
   +154 x_{+}^4+246x_{+}^3 )
   +34 (x_{-}-1)^2 x_{-}^5 (2x_{-}-3)
   \bigg)\ln (x_{-})
   \bigg]
   \bigg\} \,,
\end{eqnarray}

\begin{eqnarray}
\mathcal{C}_{\rho_D}^{\mbox{\scriptsize NLO, F}} &=& 
-4\pi^2 \bigg\{
- \frac{1}{3 (x_{-}-1) x_{-} (x_{-}-x_{+})^3 (x_{+}-1) x_{+}}
\bigg( 
   720 x_{-}^6 x_{+}^6 - 6 x_{-}^5 x_{+}^5 (80 x_{-}^2+201 x_{-}+80 x_{+}^2
   \nonumber
   \\
   &&
   +201x_{+}-1358) 
   -112 x_{+}^4 (2 x_{+}^3-9 x_{+}^2+13 x_{+}-6)  
   + x_{-}^4 x_{+}^4 (120 x_{-}^4+537 x_{-}^3+1823 x_{-}^2
   \nonumber
   \\
   &&
   -14878 x_{-}+120 x_{+}^4+537 x_{+}^3+1823 x_{+}^2-14878x_{+}+25684) 
   +x_{-}^3 x_{+}^3(-99 x_{-}^5+726 x_{-}^4
   \nonumber
   \\
   &&
   -7148 x_{-}^3+22367 x_{-}^2-27606x_{-}-99 x_{+}^5+726 x_{+}^4-7148 x_{+}^3+22367 x_{+}^2-27606 x_{+}
   \nonumber
   \\
   &&
   +19984)
   -x_{-}^2 x_{+}^2(95 x_{-}^6+1088 x_{-}^5-8406 x_{-}^4+18391 x_{-}^3-16272 x_{-}^2+7888x_{-}+95 x_{+}^6
   \nonumber
   \\
   &&
   +1088 x_{+}^5-8406 x_{+}^4+18391 x_{+}^3-16272 x_{+}^2+7888 x_{+}-2688)
   +x_{-} x_{+}(74 x_{-}^7+545 x_{-}^6
   \nonumber
   \\
   &&
   -3699 x_{-}^5+6136 x_{-}^4-2944 x_{-}^3-96x_{-}^2+x_{+}^2 (74 x_{+}^5+545 x_{+}^4-3699 x_{+}^3+6136 x_{+}^2
   \nonumber
   \\
   &&
   -2944 x_{+}-96) )
   -112 x_{-}^4 (2 x_{-}^3-9 x_{-}^2+13 x_{-}-6)
   \bigg)
   \nonumber
   \\
   &&
   - \frac{32}{3 x_{-}^2 (x_{-}-x_{+}) x_{+}^2} 
   \bigg(
   7(3-2 x_{-}) x_{-}^4 + x_{+}^4 x_{-}^4 (27 x_{-}+27 x_{+}-92) 
   +3 x_{+}^3 x_{-}^3 (x_{-}^3-10 x_{-}^2
   \nonumber
   \\
   &&
   +22 x_{-} +x_{+}^3-10 x_{+}^2+22 x_{+}-36) 
   -2 x_{+}^2 x_{-}^2 (x_{-}^4+27x_{-}^3-66 x_{-}^2+10 x_{-}+x_{+}^4+27 x_{+}^3
   \nonumber
   \\
   &&
   -66 x_{+}^2+10 x_{+} +3) 
   +2 x_{+} x_{-} (27 x_{-}^4-52 x_{-}^3+12 x_{-}^2 + 27 x_{+}^4-52 x_{+}^3+12x_{+}^2 ) 
   \nonumber
   \\
   &&
   +7x_{+}^4 (3-2 x_{+})
   \bigg) \ln (x_{+}-x_{-})
   \nonumber
   \\
   &&
   -\frac{2}{3 (x_{-}-x_{+})^3}
   \bigg( 
   -202 x_{-}^3 x_{+}^3 (3 x_{-}+3 x_{+}-16) 
   + x_{-}^2 x_{+}^2 (-627 x_{-}^3+3814 x_{-}^2-8670x_{-}
   \nonumber
   \\
   &&
   -627 x_{+}^3+3814 x_{+}^2-8670 x_{+}+11520 ) 
   + x_{+}^2(-310 x_{+}^4+1233 x_{+}^3-1920x_{+}^2+1664 x_{+} 
   \nonumber
   \\
   &&
   -768)
   + x_{-} x_{+}(465 x_{-}^5-1280 x_{-}^4-243 x_{-}^3+3840x_{-}^2-5504 x_{-}+465 x_{+}^5-1280 x_{+}^4
   \nonumber
   \\
   &&
   -243 x_{+}^3+3840 x_{+}^2
   -5504 x_{+}+3072) + x_{-}^2 (-310 x_{-}^4+1233 x_{-}^3-1920 x_{-}^2
   \nonumber
   \\
   && 
   +1664 x_{-}-768)
   \bigg) 
   \ln\left(\frac{\mu }{m_b}\right)
   \nonumber
   \\
   &&
   - 8( 42 x_{-}^2 x_{+}^2 
   + x_{+} (-10 x_{+}^2+55 x_{+}-108)    
   + x_{-} x_{+} (15 x_{-}^2-94 x_{-}+15 x_{+}^2-94 x_{+}
   \nonumber
   \\
   && 
   +186) 
   + x_{-} (-10 x_{-}^2+55 x_{-}-108) + 56 ) \bigg(
   2L\bigg(1-\frac{x_{-}}{x_{+}}\bigg) + L(x_{-}) - L(x_{+})
   \bigg)
   \nonumber
\end{eqnarray}
\begin{eqnarray}
   &&
   +\frac{4}{3 (x_{-}-x_{+})}
   \bigg[
   \frac{x_{-}-1}{x_{-}^2}\bigg(
   -30 x_{-}^4 x_{+}^4
   +3x_{-}^3 x_{+}^3 (10 x_{-}^2+10 x_{-}+25 x_{+}-77) 
   \nonumber
   \\
   &&
   +28 (2 x_{+}-3) x_{+}^2 
   + x_{-}^2 x_{+}^2 (-105 x_{-}^3+264 x_{-}^2+80 x_{-}-30 x_{+}^2+269 x_{+}-499) 
   \nonumber
   \\
   &&
   +x_{-} x_{+} ( 117x_{-}^4-503 x_{-}^3+340 x_{-}^2+124 x_{-}-160 x_{+}^2+332 x_{+}-96) 
   \nonumber
   \\
   &&
   +x_{-}^2 (-38 x_{-}^3+175 x_{-}^2-128 x_{-}+12)
   \bigg) \ln (1-x_{-})
   \nonumber
   \\
   &&
   +\frac{x_{+}-1}{x_{+}^2}\bigg(
   -30 x_{-}^4 x_{+}^4 
   + 3 x_{-}^3 x_{+}^3 (10 x_{+}^2+10 x_{+}+25 x_{-}-77)
   + x_{-}^2 x_{+}^2 (-105 x_{+}^3+264 x_{+}^2
   \nonumber
   \\
   &&
   +80 x_{+}-30 x_{-}^2+269 x_{-}-499) 
   + x_{+}^2  (-38 x_{+}^3+175 x_{+}^2-128 x_{+}+12) 
   + x_{-} x_{+} (117 x_{+}^4
   \nonumber
   \\
   &&
   -503x_{+}^3+340 x_{+}^2+124 x_{+}-160 x_{-}^2+332 x_{-}-96) 
   +28 x_{-}^2 (2x_{-}-3)
   \bigg) \ln (1-x_{+})
   \bigg]
   \nonumber
   \\
   &&
   + \frac{2}{3 (x_{-}-x_{+})^4}
   \bigg[
   \frac{1}{(x_{-}-1)^2 x_{+}^2}
   \bigg(
   -360x_{-}^7 x_{+}^7 + 12 x_{-}^6 x_{+}^6 (20 x_{-}^2-5 x_{-}+20 x_{+}^2+95 x_{+}-612) 
   \nonumber
   \\
   &&
   -24x_{+}^5 (4 x_{+}^3-9 x_{+}^2-3 x_{+}+9)  
   + x_{-}^5 x_{+}^5 (-60 x_{-}^4-510 x_{-}^3+2889x_{-}^2+8952 x_{-}-60 x_{+}^4
   \nonumber
   \\
   &&
   -1110 x_{+}^3+228 x_{+}^2+25208 x_{+}-57447) 
   + x_{-}^4 x_{+}^4 (210 x_{-}^5+114 x_{-}^4-5874 x_{-}^3+5428 x_{-}^2
   \nonumber
   \\
   &&
   +35080 x_{-}+330 x_{+}^5+1782 x_{+}^4-5552x_{+}^3-31173 x_{+}^2+96064 x_{+}-82750) 
   \nonumber
   \\
   &&
   + x_{-}^3 x_{+}^3(-186 x_{-}^6+656 x_{-}^5+4072x_{-}^4-16328 x_{-}^3+12738 x_{-}^2+9520 x_{-}-555 x_{+}^6-1138 x_{+}^5
   \nonumber
   \\
   &&
   +8874 x_{+}^4+12618x_{+}^3-66224 x_{+}^2+67048 x_{+}-15056) 
   +x_{-}^2 x_{+}^2 (44 x_{-}^7-1189 x_{-}^6+1674x_{-}^5
   \nonumber
   \\
   &&
   +4711 x_{-}^4-10760 x_{-}^3+6400 x_{-}^2-888 x_{-} + 360 x_{+}^7+86x_{+}^6-5400 x_{+}^5+3103 x_{+}^4+15544 x_{+}^3
   \nonumber
   \\
   &&
   -20360 x_{+}^2+4632x_{+} ) 
   +x_{-} x_{+} (864 x_{-}^7-2720 x_{-}^6+2224 x_{-}^5+256 x_{-}^4-624 x_{-}^3 -75 x_{+}^8
   \nonumber
   \\
   &&
   +228x_{+}^7+902 x_{+}^6-2784 x_{+}^5+1168 x_{+}^4+984x_{+}^3 ) 
   -112 (x_{-}-1)^2 x_{-}^5 (2x_{-}-3)
   \bigg) \ln (x_{-})
   \nonumber
   \\
   &&
   +\frac{1}{x_{-}^2 (x_{+}-1)^2}
   \bigg(
   360 x_{-}^7 x_{+}^7
   -12 x_{-}^6 x_{+}^6 (20 x_{-}^2+95 x_{-}+20 x_{+}^2-5 x_{+}-612) 
   \nonumber
   \\
   &&
   +112 (x_{+}-1)^2 (2 x_{+}-3)x_{+}^5 
   + x_{-}^5 x_{+}^5 (60 x_{-}^4+1110 x_{-}^3-228 x_{-}^2-25208 x_{-}+60 x_{+}^4+510x_{+}^3
   \nonumber
   \\
   &&
   -2889 x_{+}^2 -8952 x_{+}+57447) 
   -x_{-}^4 x_{+}^4(330 x_{-}^5+1782 x_{-}^4-5552x_{-}^3-31173 x_{-}^2+96064 x_{-}
   \nonumber
   \\
   &&
   +210 x_{+}^5
   +114 x_{+}^4-5874 x_{+}^3+5428 x_{+}^2+35080x_{+}-82750) 
   +x_{-}^3 x_{+}^3 (555 x_{-}^6+1138 x_{-}^5-8874 x_{-}^4
   \nonumber
   \\
   &&
   -12618 x_{-}^3
   +66224x_{-}^2-67048 x_{-} + 186 x_{+}^6-656 x_{+}^5-4072 x_{+}^4+16328 x_{+}^3-12738 x_{+}^2-9520 x_{+}
   \nonumber
   \\
   &&
   +15056 ) 
   -x_{-}^2 x_{+}^2 ( 360 x_{-}^7+86 x_{-}^6-5400 x_{-}^5+3103x_{-}^4+15544 x_{-}^3-20360 x_{-}^2+4632 x_{-} + 44 x_{+}^7
   \nonumber
   \\
   &&
   -1189 x_{+}^6
   +1674x_{+}^5+4711 x_{+}^4-10760 x_{+}^3+6400 x_{+}^2-888x_{+} ) 
   +x_{-} x_{+} ( 75 x_{-}^8-228x_{-}^7-902 x_{-}^6
   \nonumber
   \\
   &&
   +2784 x_{-}^5
   -1168 x_{-}^4-984 x_{-}^3 -864 x_{+}^7+2720 x_{+}^6-2224 x_{+}^5-256 x_{+}^4+624 x_{+}^3 ) 
   \nonumber
   \\
   &&
   +24 x_{-}^5 (4 x_{-}^3-9 x_{-}^2-3x_{-}+9)
   \bigg) \ln (x_{+})
   \bigg]
   \bigg\}\,,
\end{eqnarray}

\section{Darwin coefficients for the moments with low cut}
\label{App:Mnrcut}

Here we present analytical expressions for the Darwin coefficients
of the moments $M_n(r_{cut})$ defined in Eqs.~(\ref{momdef}) and (\ref{momdefLONLO}),
which depend on the low 
cut $r_{\rm cut}$ in the lepton pair invariant mass squared. We define
\begin{eqnarray}
 t_0 &=& \frac{1}{2}\ln\left(\frac{x_{+}(r_{cut})}{x_{-}(r_{cut})}\right)\,. 
\end{eqnarray}
At LO they read

\begin{eqnarray}
 M_{0,\rho_D}^{\mbox{\scriptsize LO}}(r_{cut}) &=& 
  -\frac{4}{3} (5 \rho +21) \rho ^{3/2} \sinh (3 t_0)+8 \left(3 \rho ^2+4\right) t_0+\frac{10}{3} \rho ^2 \sinh (4t_0)
   \\
   &&
   -\frac{4}{3} \left(33 \rho ^2+153 \rho +104\right) \sqrt{\rho } \sinh (t_0)+16 \left(5 \rho ^2+10 \rho
   +1\right) \coth (t_0)
   \nonumber
   \\
   &&
   -16 \left(\rho ^2+10 \rho +5\right) \sqrt{\rho } \text{csch}(t_0)+\frac{40}{3} (5 \rho +6)
   \rho  \sinh (2 t_0)\,,
   \nonumber
   \\
 &&
 \nonumber
   \\
   M_{1,\rho_D}^{\mbox{\scriptsize LO}}(r_{cut}) &=& 
 -\frac{8}{3} \rho ^{5/2} \sinh (5 t_0)+\frac{1}{3} (25 \rho +73) \rho ^2 \sinh (4 t_0)
   \\
   &&
 +\frac{8}{3} \left(37 \rho
   ^2+109 \rho +56\right) \rho  \sinh (2 t_0)-\frac{4}{9} \left(15 \rho ^2+188 \rho +183\right) \rho ^{3/2} \sinh (3
   t_0)
      \nonumber
   \\
   &&
   +\frac{4}{3} \left(27 \rho ^3-69 \rho ^2+8 \rho +24\right) t_0-\frac{4}{3} \left(33 \rho ^3+322 \rho ^2+377
   \rho +152\right) \sqrt{\rho } \sinh (t_0)
   \nonumber
   \\
   &&
   +16 \left(7 \rho ^3+35 \rho ^2+21 \rho +1\right) \coth (t_0)-16
   \left(\rho ^3+21 \rho ^2+35 \rho +7\right) \sqrt{\rho } \text{csch}(t_0)\,,
   \nonumber
   \\
 &&
 \nonumber
   \\
   M_{2,\rho_D}^{\mbox{\scriptsize LO}}(r_{cut}) &=& 
 -\frac{4}{15} (35 \rho +83) \rho ^{5/2} \sinh (5 t_0)+\frac{20}{9} \rho ^3 \sinh (6 t_0)
   \\
   &&
 +\frac{8}{3} \left(5 \rho
   ^2+37 \rho +32\right) \rho ^2 \sinh (4 t_0)
   \nonumber
   \\
   &&
   +\frac{4}{3} \left(98 \rho ^3+547 \rho ^2+610 \rho +188\right) \rho  \sinh (2
   t_0)
      \nonumber
   \\
   &&
   -\frac{4}{9} \left(15 \rho ^3+376 \rho ^2+880 \rho +407\right) \rho ^{3/2} \sinh (3 t_0)
      \nonumber
   \\
   &&
   +\frac{16}{3} \left(9\rho ^4-56 \rho ^3-131 \rho ^2+4 \rho +6\right) t_0
      \nonumber
   \\
   &&
   -\frac{4}{3} \left(33 \rho ^4+557 \rho ^3+997 \rho ^2+769 \rho
   +200\right) \sqrt{\rho } \sinh (t_0)
      \nonumber
   \\
   &&
   +16 \left(9 \rho ^4+84 \rho ^3+126 \rho ^2+36 \rho +1\right) \coth (t_0)
   \nonumber
   \\
   &&
   -16\left(\rho ^4+36 \rho ^3+126 \rho ^2+84 \rho +9\right) \sqrt{\rho } \text{csch}(t_0)\,,
   \nonumber
\end{eqnarray}

\begin{eqnarray}
M_{3,\rho_D}^{\mbox{\scriptsize LO}}(r_{cut}) &=& -\frac{40}{21} \rho ^{7/2} \sinh (7 t_0)+\frac{2}{3} (15 \rho +31) \rho ^3 \sinh (6 t_0)
   \\
   &&
 -\frac{4}{15} \left(75 \rho^2+424 \rho +339\right) \rho ^{5/2} \sinh (5 t_0)
    \nonumber
   \\
   &&
 +\frac{1}{3} \left(55 \rho ^3+747 \rho ^2+1515 \rho +663\right) \rho ^2 \sinh (4 t_0)
    \nonumber
   \\
   &&
   -\frac{4}{3} \left(5 \rho ^4+209 \rho ^3+882 \rho ^2+921 \rho +261\right) \rho ^{3/2} \sinh (3 t_0)
      \nonumber
   \\
   &&
   +4
   \left(15 \rho ^5-165 \rho ^4-925 \rho ^3-585 \rho ^2+8 \rho +8\right) t_0
      \nonumber
   \\
   &&
   +\frac{2}{3} \sqrt{\rho } \bigg(\sqrt{\rho }
   \left(244 \rho ^4+2223 \rho ^3+4191 \rho ^2+2772 \rho +576\right) \sinh (2 t_0)
      \nonumber
   \\
   &&
   -2 \left(33 \rho ^5+858 \rho ^4+2200 \rho
   ^3+1938 \rho ^2+1377 \rho +248\right) \sinh (t_0)\bigg)
         \nonumber
   \\
   &&
   +16 \left(11 \rho ^5+165 \rho ^4+462 \rho ^3+330 \rho ^2+55 \rho
   +1\right) \coth (t_0)
         \nonumber
   \\
   &&
   -16 \left(\rho ^5+55 \rho ^4+330 \rho ^3+462 \rho ^2+165 \rho +11\right) \sqrt{\rho }
   \text{csch}(t_0)\,,
   \nonumber
   \\
   &&
   \nonumber
\\
 M_{4,\rho_D}^{\mbox{\scriptsize LO}}(r_{cut}) &=& 
 -\frac{4}{21} (55 \rho +103) \rho ^{7/2} \sinh (7 t_0)+\frac{5}{3} \rho ^4 \sinh (8 t_0)
    \\
    &&
 +\frac{8}{9} \left(30 \rho
   ^2+143 \rho +108\right) \rho ^3 \sinh (6 t_0)
       \nonumber
    \\
    &&
   -\frac{4}{15} \left(130 \rho ^3+1291 \rho ^2+2371 \rho +1002\right) \rho^{5/2} \sinh (5 t_0)
       \nonumber
    \\
    &&
   +\frac{2}{3} \left(35 \rho ^4+752 \rho ^3+2666 \rho ^2+2640 \rho +723\right) \rho ^2 \sinh (4t_0)
       \nonumber
    \\
    &&
   +\frac{8}{3} \left(73 \rho ^5+988 \rho ^4+2815 \rho ^3+2916 \rho ^2+1377 \rho +206\right) \rho  \sinh (2t_0)
       \nonumber
    \\
    &&
   -\frac{4}{9} \left(15 \rho ^5+941 \rho ^4+6243 \rho ^3+11115 \rho ^2+6981 \rho +1359\right) \rho ^{3/2} \sinh (3t_0)
    \nonumber
    \\
    &&
   +\frac{8}{3} \left(27 \rho ^6-456 \rho ^5-4495 \rho ^4-6840 \rho ^3-2157 \rho ^2+16 \rho +12\right)t_0
       \nonumber
    \\
    &&
   -\frac{4}{3} (33 \rho ^6+1225 \rho ^5+4291 \rho ^4+2779 \rho ^3+2577 \rho ^2+2249 \rho 
       \nonumber
    \\
    &&
   +296) \sqrt{\rho }
   \sinh (t_0)
    \nonumber
    \\
    &&
   +16 \left(13 \rho ^6+286 \rho ^5+1287 \rho ^4+1716 \rho ^3+715 \rho ^2+78 \rho +1\right) \coth (t_0)
    \nonumber
    \\
    &&
   -16\left(\rho ^6+78 \rho ^5+715 \rho ^4+1716 \rho ^3+1287 \rho ^2+286 \rho +13\right) \sqrt{\rho } \text{csch}(t_0)\,,
    \nonumber
\end{eqnarray}
at NLO the expressions are very lengthy and are provided in the file ``Coef.m''.

\section{Darwin coefficients for moments}
\label{App:Mn0}

The moments integrated over the whole phase space can be obtained from the results of the previous section in the 
special case $r_{\rm cut}=0$. The zeroth moment, or total width reads

\begin{eqnarray}
 M_{0,\rho_D}^{\mbox{\scriptsize LO}}(0) &=& C_{\rho_D}^{\mbox{\scriptsize LO}} 
 = \frac{1}{3} \left(5 \rho ^4+8 \rho ^3-24 \rho ^2-12 \left(3 \rho ^2+4\right) \ln (\rho )+88 \rho -77\right)\,,
 \label{CrhoDtotLO}
 \\
 &&
 \nonumber
 \\
 M_{0,\rho_D}^{\mbox{\scriptsize NLO, F}}(0) &=& C_{\rho_D}^{\mbox{\scriptsize NLO, F}} 
 \nonumber
 \\
 &=& 
 \frac{64}{9} \left(\rho ^4-2 \rho ^3-3 \rho ^2+14 \rho -6 \ln (\rho )-10\right) \ln \left(\frac{\mu}{m_b}\right)
 \nonumber
 \\
 &&
 +\frac{64}{9} \sqrt{\rho } \left(27 \rho ^2+16 \rho +13\right) \Li_2\left(1-\sqrt{\rho}\right)
  \nonumber
 \\
 &&
 +\frac{1}{9} \left(-432 \rho ^{5/2}-256 \rho ^{3/2}+45 \rho ^4+24 \rho ^3-744 \rho ^2+768 \rho -208 \sqrt{\rho}-147\right) \Li_2(1-\rho )
 \nonumber
 \\
 &&
   +\frac{1}{18} \rho  \left(15 \rho ^3-16 \rho ^2-500 \rho -96\right) \ln ^2(\rho)
 \nonumber
 \\
 &&
   +\frac{1}{270} \left(255 \rho ^4-8 \rho ^3-23390 \rho ^2+780 \rho -6600\right) \ln (\rho )
 \nonumber
 \\
 &&
   +\frac{1}{9} \left(15 \rho ^4+20\rho ^3+256 \rho ^2+468 \rho +83\right) \ln (1-\rho ) \ln (\rho )
 \nonumber
 \\
 &&
   +\frac{4305 \rho ^4+56740 \rho ^3+44072 \rho ^2-61044 \rho-41721}{1080}
 \nonumber
 \\
 &&
   -\frac{\left(255 \rho ^5+3232 \rho ^4+9580 \rho ^3-5640 \rho ^2-5815 \rho -2200 \right) \ln (1-\rho )}{270 \rho}
 \nonumber
 \\
 &&
 - \frac{98}{45} \bigg( \frac{1}{\rho } +\frac{\ln (1-\rho )}{\rho ^2} \bigg)\,,
 \label{CrhoDCFtotNLO}
 \\
 &&
 \nonumber
 \\
 M_{0,\rho_D}^{\mbox{\scriptsize NLO, A}}(0) &=& C_{\rho_D}^{\mbox{\scriptsize NLO, A}} 
 \nonumber
\\
 &=&
 \frac{2}{3} \left(9 \rho ^4-28 \rho ^3+18 \rho ^2+36 \rho -24 \ln (\rho )-35\right) \ln \left(\frac{\mu}{m_b}\right)
\nonumber
\\
&&
+\frac{8}{9} \sqrt{\rho } \left(57 \rho ^2+50 \rho -35\right) \Li_2\left(1-\sqrt{\rho}\right)
\nonumber
\\
&&
-\frac{2}{9} \left(57 \rho ^{5/2}+50 \rho ^{3/2}+69 \rho ^3+37 \rho ^2+108 \rho -35 \sqrt{\rho }-24\right)\Li_2(1-\rho )
\nonumber
\\
&&
   +\frac{1}{9} \left(-17 \rho ^3-23 \rho ^2+45 \rho +39\right) \ln ^2(\rho )
\nonumber
\\
&&
   -\frac{2}{9} \left(26 \rho^3+18 \rho ^2+75 \rho -1\right) \ln (1-\rho ) \ln (\rho )
   \nonumber
\\
&&
   +\frac{1}{270} \left(420 \rho ^4-5117 \rho ^3+1455 \rho ^2-4290
   \rho +6930\right) \ln (\rho )
   \nonumber
\\
&&
   -\frac{1480 \rho ^4-5285 \rho ^3+701 \rho ^2+22203 \rho-19813}{540}
\nonumber
\\
&&
  -\frac{\left(420 \rho ^5-4097 \rho ^4+2295 \rho ^3-1740 \rho ^2+4160 \rho -1395 \right) \ln (1-\rho )}{270 \rho}
  \nonumber
\\
&&
- \frac{119}{90}\bigg( \frac{1}{\rho} + \frac{\ln (1-\rho )}{\rho^2} \bigg)\,.
\label{CrhoDCAtotNLO}
\end{eqnarray}
Note that there is no $1/\rho$ singularity at small $\rho$. The same applies to higher moments. 
The first moment reads
\begin{eqnarray}
 M_{1,\rho_D}^{\mbox{\scriptsize LO}}(0) &=& \frac{1}{18} \bigg(
 -729 + 1831 \rho - 1296 \rho^2 + 144 \rho^3 + 
    41 \rho^4 + 9 \rho^5 
    \nonumber
    \\
    &&
    - 12 (24 + 8 \rho - 69 \rho^2 + 27 \rho^3) \ln(\rho)
    \bigg)\,,
    \\
    &&
    \nonumber
    \\
M_{1,\rho_D}^{\mbox{\scriptsize NLO, F}}(0) &=& 
   \frac{32}{135} (9 \rho ^5-25 \rho ^4-180 \rho ^3-540 \rho ^2+60 \left(12 \rho ^2-\rho -3\right) \ln (\rho )+1195 \rho
    \nonumber
    \\
    &&
    -459) \ln \left(\frac{\mu }{m_b}\right)
   +\frac{64}{9} \sqrt{\rho } \left(27 \rho ^3+166 \rho ^2-145 \rho+40\right) \Li_2\left(1-\sqrt{\rho }\right)
    \nonumber
    \\
    &&
   +\frac{1}{18} (-864 \rho ^{7/2}-5312 \rho ^{5/2}+4640 \rho ^{3/2}+27
   \rho ^5+103 \rho ^4-3952 \rho ^3+2832 \rho ^2
    \nonumber
    \\
    &&
   +3303 \rho -1280 \sqrt{\rho }-485) \Li_2(1-\rho )
   +\frac{1}{36} \rho (9 \rho ^4+45 \rho ^3-2500 \rho ^2+92 \rho 
    \nonumber
    \\
    &&
    -640) \ln ^2(\rho )
   +\frac{1}{540} (291 \rho ^5+1218 \rho^4-142110 \rho ^3-19510 \rho ^2-12545 \rho 
     \nonumber
    \\
    &&
    -21120) \ln (\rho )
   +\frac{1}{18} (9 \rho ^5+29 \rho ^4+444 \rho
   ^3+2108 \rho ^2+1933 \rho 
       \nonumber
    \\
    &&
    +121) \ln (1-\rho ) \ln (\rho )
   +\frac{1}{3240 \rho }(4077 \rho ^6+126439 \rho ^5+1536734 \rho ^4
    \nonumber
    \\
    &&
    -1850526
   \rho ^3+561605 \rho ^2-376649 \rho -1680)
   -\frac{1}{540 \rho ^2}(291 \rho ^7+4123 \rho ^6
      \nonumber
    \\
    && 
   +23200 \rho ^5+62960 \rho ^4-75035\rho ^3-14371 \rho ^2-1448 \rho +280) \ln (1-\rho )\,,
    \\
    &&
    \nonumber
    \\
M_{1,\rho_D}^{\mbox{\scriptsize NLO, A}}(0) &=& 
 \frac{1}{45} (81 \rho ^5-425 \rho ^4+480 \rho ^3-3060 \rho ^2+60 \left(33 \rho ^2+\rho -12\right) \ln (\rho )+4655 \rho
    \nonumber
    \\
    &&
    -1731) \ln \left(\frac{\mu }{m_b}\right)
   +\frac{8}{9} \sqrt{\rho } \left(57 \rho ^3+13 \rho ^2+435 \rho
   -89\right) \Li_2\left(1-\sqrt{\rho }\right)
   \nonumber
    \\
    &&
   -\frac{1}{9} (114 \rho ^{7/2}+26 \rho ^{5/2}+870 \rho ^{3/2}+97 \rho^4-2 \rho ^3+1028 \rho ^2+526 \rho -178 \sqrt{\rho }
    \nonumber
    \\
    &&
    -89) \Li_2(1-\rho )
   +\frac{1}{18} \left(-35 \rho ^4+44 \rho^3-337 \rho ^2+174 \rho +78\right) \ln ^2(\rho )
    \nonumber
    \\
    &&
   -\frac{1}{9} \left(31 \rho ^4+70 \rho ^3+312 \rho ^2+326 \rho -11\right)
   \ln (1-\rho ) \ln (\rho )
      \nonumber
    \\
    &&
   +\frac{1}{270} \left(126 \rho ^5-3719 \rho ^4+6350 \rho ^3-47625 \rho ^2-11260 \rho +8910\right)
   \ln (\rho )
       \nonumber
    \\
    &&
   -\frac{666 \rho ^6-4767 \rho ^5+13987 \rho ^4-140085 \rho ^3+178068 \rho ^2-48124 \rho +255}{810 \rho}
   \nonumber
    \\
    &&
   -\frac{1}{270 \rho ^2}(126 \rho ^7-3089 \rho ^6+1630 \rho ^5-14925 \rho ^4+14470 \rho ^3+2249 \rho ^2
    \nonumber
    \\
    &&
    -546 \rho +85) \ln (1-\rho)\,. 
\end{eqnarray}
The second moment reads

\begin{eqnarray}
 M_{2,\rho_D}^{\mbox{\scriptsize LO}}(0) &=& \frac{2}{45} \bigg(
 -1119 + 4420 \rho + 5275 \rho^2 - 9200 \rho^3 + 
    575 \rho^4 + 44 \rho^5 + 5 \rho^6 
    \nonumber
    \\
    &&
    - 60 (6 + 4 \rho - 131 \rho^2 - 56 \rho^3 + 9 \rho^4) \ln(\rho)
       \bigg)\,,
       \\
       &&
       \nonumber
 \\
 M_{2,\rho_D}^{\mbox{\scriptsize NLO, F}}(0) &=& 
 \frac{32}{135} (4 \rho ^6-14 \rho ^5-275 \rho ^4-4600 \rho ^3+3200 \rho ^2+60 (40 \rho ^3+70 \rho ^2
       \nonumber
    \\
    &&
   -2 \rho -3)
   \ln (\rho )+2246 \rho -561) \ln \left(\frac{\mu }{m_b}\right)
    \nonumber
    \\
    &&
   +\frac{64}{9} \sqrt{\rho } \left(27 \rho ^4+424
   \rho ^3-590 \rho ^2-600 \rho +67\right) \Li_2\left(1-\sqrt{\rho }\right)
    \nonumber
    \\
    &&
   +\frac{2}{9} (-216 \rho ^{9/2}-3392 \rho^{7/2}+4720 \rho ^{5/2}+4800 \rho ^{3/2}+3 \rho ^6+18 \rho ^5-1500 \rho ^4
    \nonumber
    \\
    &&
   -752 \rho ^3
   +9989 \rho ^2+1394 \rho -536 \sqrt{\rho}-150) \Li_2(1-\rho )
    \nonumber
    \\
    &&
   +\frac{1}{9} \rho  \left(\rho ^5+6 \rho ^4-908 \rho ^3-542 \rho ^2+2550 \rho -296\right)\ln ^2(\rho )
    \nonumber
    \\
    &&
   +\frac{1}{135} (44 \rho ^6+391 \rho ^5-65765 \rho ^4-62046 \rho ^3+121845 \rho ^2-7190 \rho 
   \nonumber
    \\
    &&
    -6522) \ln (\rho )
   +\frac{2}{9} (\rho ^6+6 \rho ^5+100 \rho ^4+704 \rho ^3+1995 \rho ^2+798 \rho +22)
    \nonumber
    \\
    &&
    \times \ln (1-\rho ) \ln(\rho )
   + \frac{1}{8100 \rho }(4720 \rho ^7+235363 \rho ^6+9045850 \rho ^5-9532420 \rho ^4
    \nonumber
    \\
    &&
    -2786860 \rho ^3+4462265 \rho ^2-1427358 \rho-1560)
   -\frac{1}{135 \rho ^2}(44 \rho ^8+858 \rho ^7
    \nonumber
    \\
    &&
    +5055 \rho ^6+39924 \rho ^5+6450 \rho ^4-48794 \rho ^3-3383 \rho ^2-180
   \rho +26) \ln (1-\rho )\,,
    \nonumber
    \\
    &&
\end{eqnarray}

\begin{eqnarray}
 M_{2,\rho_D}^{\mbox{\scriptsize NLO, A}}(0) &=& 
 \frac{2}{45} (18 \rho ^6-143 \rho ^5+250 \rho ^4-7900 \rho ^3+3950 \rho ^2+60 (50 \rho ^3+130 \rho ^2+\rho 
     \nonumber
    \\
    && 
  -6)
   \ln (\rho )+4907 \rho -1082) \ln \left(\frac{\mu }{m_b}\right)
    \nonumber
    \\
    &&   
   +\frac{8}{45} \sqrt{\rho } \left(285 \rho ^4-780
   \rho ^3+10486 \rho ^2+6980 \rho -715\right) \Li_2\left(1-\sqrt{\rho }\right)
     \nonumber
    \\
    &&
   -\frac{1}{45} (570 \rho ^{9/2}-1560
   \rho ^{7/2}+20972 \rho ^{5/2}+13960 \rho ^{3/2}+339 \rho ^5-125 \rho ^4
       \nonumber
    \\
    && 
   +8810 \rho ^3+34470 \rho ^2+4385 \rho -1430 \sqrt{\rho}-531) \Li_2(1-\rho )
       \nonumber
    \\
    &&
   +\frac{1}{90} \left(-113 \rho ^5+295 \rho ^4-3160 \rho ^3-11180 \rho ^2+1410 \rho +390\right)
   \ln ^2(\rho )
       \nonumber
    \\
    &&
   -\frac{1}{45} \left(113 \rho ^5+465 \rho ^4+2790 \rho ^3+8850 \rho ^2+2695 \rho -77\right) \ln (1-\rho ) \ln(\rho )
    \nonumber
    \\
    &&
   +\frac{1}{9450}(1960 \rho ^6-107302 \rho ^5+485730 \rho ^4-4731475 \rho ^3-9935345 \rho ^2
    \nonumber
    \\
    &&   
   -756140 \rho +355320) \ln (\rho )
   -\frac{1}{113400 \rho }(41440 \rho ^7-699853 \rho ^6+10454561 \rho ^5
       \nonumber
    \\
    &&
   -118412512 \rho ^4+60323932 \rho ^3+57175009 \rho ^2-8895837
   \rho +13260)
    \nonumber
    \\
    &&
   -\frac{1}{9450 \rho ^2}(1960 \rho ^8-90712 \rho ^7+30870 \rho ^6-1464855 \rho ^5+102375 \rho ^4
    \nonumber
    \\
    &&   
   +1384040 \rho^3+45962 \rho ^2-10745 \rho +1105) \ln (1-\rho )\,.
    \nonumber
    \\
    &&
\end{eqnarray}
The third moment reads

\begin{eqnarray}
 M_{3,\rho_D}^{\mbox{\scriptsize LO}}(0) &=& \frac{1}{210} \bigg(-11897 + 66843 \rho + 377475 \rho^2 - 207025 \rho^3 - 
     235375 \rho^4 + 9597 \rho^5 
     \nonumber
     \\
     &&
     + 357 \rho^6 + 25 \rho^7\bigg)
     - 2 (8 + 8 \rho - 585 \rho^2 - 925 \rho^3 - 165 \rho^4 + 15 \rho^5) \ln(\rho)\,,
\end{eqnarray}

\begin{eqnarray}
 M_{3,\rho_D}^{\mbox{\scriptsize NLO, F}}(0) &=& 
 \frac{32}{315} (5 \rho ^7-21 \rho ^6-861 \rho ^5-32375 \rho ^4-22925 \rho ^3+49245 \rho ^2+420 (30 \rho ^4
    \nonumber
   \\
   &&
 +125 \rho
   ^3+75 \rho ^2-\rho -1) \ln (\rho )+8421 \rho -1489) \ln \left(\frac{\mu }{m_b}\right)
    \nonumber
    \\
    &&
   +\frac{64}{315}
   \sqrt{\rho } (945 \rho ^5+27650 \rho ^4-44742 \rho ^3-206724 \rho ^2-52675 \rho 
       \nonumber
   \\
   &&
   +3290)
   \Li_2\left(1-\sqrt{\rho }\right)
   +\frac{1}{630} (-30240 \rho ^{11/2}-884800 \rho ^{9/2}+1431744 \rho ^{7/2}
       \nonumber
   \\
   &&
   +6615168
   \rho ^{5/2}+1685600 \rho ^{3/2}+225 \rho ^7+2163 \rho ^6-282849 \rho ^5-732795 \rho ^4
       \nonumber
   \\
   &&
   +5696425 \rho ^3+5460875 \rho ^2+294903
   \rho -105280 \sqrt{\rho }-23379) \Li_2(1-\rho )
       \nonumber
    \\
    &&
   +\frac{1}{1260}\rho  (75 \rho ^6+721 \rho ^5-168623 \rho ^4-294105
   \rho ^3+2277520 \rho ^2+1440740 \rho 
    \nonumber
    \\
    &&
   -63840) \ln ^2(\rho )
   + \frac{1}{132300}(28125 \rho ^7+408534 \rho ^6-98525252
   \rho ^5
    \nonumber
    \\
    &&
   -237554415 \rho ^4+671252260 \rho ^3+615302485 \rho ^2-9618105 \rho -7279440) \ln (\rho )
     \nonumber
    \\
    &&
    +\frac{1}{630}
   (75 \rho ^7+721 \rho ^6+12817 \rho ^5+89355 \rho ^4+641095 \rho ^3+791525 \rho ^2
    \nonumber
    \\
    &&
   +165501 \rho +2287) \ln (1-\rho )\ln (\rho )
    \nonumber
    \\
    &&
   + \frac{1}{1587600 \rho}(513975 \rho ^8+38403937 \rho ^7+3013994353 \rho ^6-1396552545 \rho ^5
       \nonumber
    \\
    &&
   -12219126399 \rho ^4+9100014071 \rho^3+1823764119 \rho ^2-360870391 \rho -141120)
    \nonumber
    \\
    &&
   -\frac{1}{132300\rho ^2}
   (28125 \rho ^9+747999 \rho ^8+3631243 \rho ^7+51208185
   \rho ^6+143993115 \rho ^5
       \nonumber
    \\
    &&
   -105515375 \rho ^4-90995499 \rho ^3-3003993 \rho ^2-105560 \rho +11760) \ln (1-\rho )\,,
    \nonumber
    \\
    &&
\end{eqnarray}

\begin{eqnarray}
 M_{3,\rho_D}^{\mbox{\scriptsize NLO, A}}(0) &=& 
 \frac{1}{105} (45 \rho ^7-504 \rho ^6+1281 \rho ^5-96250 \rho ^4-115325 \rho ^3+178080 \rho ^2+420 (70 \rho ^4
     \nonumber
    \\
    &&
 +425 \rho
   ^3+300 \rho ^2+\rho -4) \ln (\rho )+38479 \rho -5806) \ln \left(\frac{\mu }{m_b}\right)
    \nonumber
    \\
    &&
   +\frac{8}{45}\sqrt{\rho } \left(285 \rho ^5-2285 \rho ^4+26034 \rho ^3+67734 \rho ^2+16225 \rho -985\right) \Li_2\left(1-\sqrt{\rho}\right)
    \nonumber
    \\
    &&
   -\frac{1}{90} (1140 \rho ^{11/2}-9140 \rho ^{9/2}+104136 \rho ^{7/2}+270936 \rho ^{5/2}+64900 \rho ^{3/2}
    \nonumber
    \\
    &&
   +537
   \rho ^6-718 \rho ^5+19905 \rho ^4+291560 \rho ^3+238785 \rho ^2+13002 \rho -3940 \sqrt{\rho }
       \nonumber
    \\
    &&
   -1143) \Li_2(1-\rho)
   -\frac{1}{180} (179 \rho ^6-926 \rho ^5+6555 \rho ^4+115040 \rho ^3+81120 \rho ^2
       \nonumber
    \\
    &&
   -4140 \rho -780) \ln ^2(\rho)
   -\frac{1}{90} (179 \rho ^6+1134 \rho ^5+8655 \rho ^4+60560 \rho ^3
    \nonumber
    \\
    &&
   +55575 \rho ^2+7974 \rho -181) \ln (1-\rho )\ln (\rho )
   +\frac{1}{18900}(2100 \rho ^7-185481 \rho ^6
      \nonumber
    \\
    &&
    +1615714 \rho ^5
   -18193665 \rho ^4-99214640 \rho ^3-68413730 \rho^2
   -2551920 \rho 
    \nonumber
    \\
    &&
   +774060) \ln (\rho )
   -\frac{1}{226800 \rho }
   (44400 \rho ^8-1291087 \rho ^7+43744586 \rho ^6
    \nonumber
    \\
    &&
   -591368493 \rho^5-675628320 \rho ^4+1038554743 \rho ^3+207381654 \rho ^2
    \nonumber
    \\
    &&
   -21449723 \rho +12240)
   -\frac{1}{6300 \rho ^2}(700 \rho ^9-52867
   \rho ^8+1158 \rho ^7-1969695 \rho ^6
    \nonumber
    \\
    &&
   -3476340 \rho ^5+3752175 \rho ^4+1733242 \rho ^3+15967 \rho ^2-4680 \rho +340) \ln (1-\rho )\,.
    \nonumber
    \\
    &&
\end{eqnarray}
Finally, the fourth moment reads

\begin{eqnarray}
 M_{4,\rho_D}^{\mbox{\scriptsize LO}}(0) &=& \frac{1}{630} \bigg(-39213 + 295376 \rho + 3814356 \rho^2 + 1916880 \rho^3 - 4547200 \rho^4 
    \nonumber
    \\
    &&
    - 1484112 \rho^5 + 42924 \rho^6 + 
    944 \rho^7 + 45 \rho^8 - 
    840 (12 + 16 \rho - 2157 \rho^2 
    \nonumber
    \\
    &&
    - 6840 \rho^3 - 4495 \rho^4 - 456 \rho^5 + 27 \rho^6) \ln(\rho)
       \bigg)\,,
\end{eqnarray}

\begin{eqnarray}
 M_{4,\rho_D}^{\mbox{\scriptsize NLO, F}}(0) &=& 
 \frac{32}{945} (9 \rho ^8-44 \rho ^7-3234 \rho ^6-214620 \rho ^5-568400 \rho ^4+268716 \rho ^3+485394 \rho ^2
     \nonumber
    \\
    &&
 +420 (168
   \rho ^5+1225 \rho ^4+1764 \rho ^3+546 \rho ^2-4 \rho -3) \ln (\rho )+37084 \rho 
    \nonumber
    \\
    &&
   -4905) \ln \left(\frac{\mu}{m_b}\right)
   +\frac{64}{315} \sqrt{\rho } (945 \rho ^6+44240 \rho ^5-69919 \rho ^4-925752 \rho ^3
   \nonumber
    \\
    &&
    -832349 \rho^2 -105560 \rho +4235) \Li_2\left(1-\sqrt{\rho }\right)
   +\frac{1}{630} (-30240 \rho ^{13/2}
    \nonumber
    \\
    &&
   -1415680 \rho
   ^{11/2}+2237408 \rho ^{9/2}+29624064 \rho ^{7/2}+26635168 \rho ^{5/2}+3377920 \rho ^{3/2}
       \nonumber
    \\
    &&
   +135 \rho ^8+1896 \rho ^7-355152 \rho
   ^6-1979544 \rho ^5+15640520 \rho ^4+37972200 \rho ^3
       \nonumber
    \\
    &&
   +14596736 \rho ^2+417608 \rho -135520 \sqrt{\rho }-25089)
   \Li_2(1-\rho )
    \nonumber
    \\
    &&
   +\frac{1}{1260}
   \rho (45 \rho ^7+632 \rho ^6-209944 \rho ^5-723128 \rho ^4+7807380 \rho ^3+14560000 \rho^2
          \nonumber
    \\
    &&
   +3845240 \rho -89600) \ln ^2(\rho )
   +\frac{1}{793800}
   (116685 \rho ^8+2469368 \rho ^7-814929696 \rho ^6
          \nonumber
    \\
    &&
   -3753287832
   \rho ^5+12993499260 \rho ^4+31919295240 \rho ^3+11621886360 \rho ^2
          \nonumber
    \\
    &&
   -61408200 \rho -47864880) \ln (\rho)
   +\frac{1}{630} (45 \rho ^8+632 \rho ^7+11816 \rho ^6+27832 \rho ^5
    \nonumber
    \\
    &&
    +982940 \rho ^4+2898840 \rho ^3+1823192 \rho^2+229176 \rho +1717) \ln (1-\rho ) \ln (\rho )
    \nonumber
    \\
    &&
   +\frac{1}{19051200 \rho }(3824955 \rho ^9+398946984 \rho ^8+53438905548 \rho^7+31280266104 \rho ^6
       \nonumber
    \\
    &&
   -690322598704 \rho ^5+34934038344 \rho ^4+537272304948 \rho ^3+38170138072 \rho ^2
       \nonumber
    \\
    &&
   -5174932491 \rho
   -893760)
   -\frac{1}{793800 \rho ^2}(116685 \rho ^{10}+4096448 \rho ^9+11921784 \rho ^8
    \nonumber
    \\
    &&
   +194037648 \rho ^7+2485981680 \rho
   ^6+982489200 \rho ^5-2771006448 \rho ^4
    \nonumber
    \\
    &&
   -891056784 \rho ^3-16195773 \rho ^2-421680 \rho +37240) \ln (1-\rho )\,,
    \nonumber
    \\
    &&
\end{eqnarray}

\begin{eqnarray}
 M_{4,\rho_D}^{\mbox{\scriptsize NLO, A}}(0) &=& 
 \frac{1}{315} (81 \rho ^8-1216 \rho ^7+4284 \rho ^6-585480 \rho ^5-2172100 \rho ^4+706944 \rho ^3+1894116 \rho ^2
    \nonumber
    \\
    &&
 +840(186 \rho ^5+1975 \rho ^4+3288 \rho ^3+1122 \rho ^2+2 \rho -6) \ln (\rho )+172616 \rho 
    \nonumber
    \\
    &&
   -19245) \ln\left(\frac{\mu }{m_b}\right)
   +\frac{8}{315} \sqrt{\rho } (1995 \rho ^6-31150 \rho ^5+335545 \rho ^4+2063436 \rho^3
    \nonumber
    \\
    &&
   +1750189 \rho ^2+220290 \rho -8785) \Li_2\left(1-\sqrt{\rho }\right)
    \nonumber
    \\
    &&
   -\frac{2}{315} (1995 \rho ^{13/2}-31150
   \rho ^{11/2}+335545 \rho ^{9/2}+2063436 \rho ^{7/2}+1750189 \rho ^{5/2}
       \nonumber
    \\
    &&
   +220290 \rho ^{3/2}+780 \rho ^7-2128 \rho ^6+19943 \rho
   ^5+1425935 \rho ^4+2874095 \rho ^3
    \nonumber
    \\
    &&
   +1064175 \rho ^2+31787 \rho -8785 \sqrt{\rho }-2097) \Li_2(1-\rho)
    \nonumber
    \\
    &&
   -\frac{1}{315} (260 \rho ^7-2226 \rho ^6+2086 \rho ^5+587440 \rho ^4+1163470 \rho ^3+375060 \rho ^2
       \nonumber
    \\
    &&
   -9975 \rho-1365) \ln ^2(\rho )
   -\frac{2}{315} (260 \rho ^7+2324 \rho ^6+21231 \rho ^5+278075 \rho ^4
       \nonumber
    \\
    &&
   +529655 \rho ^3 +230475
   \rho ^2+19299 \rho -349) \ln (1-\rho ) \ln (\rho )
    \nonumber
    \\
    &&
   +\frac{1}{396900}(26460 \rho ^8-3456445 \rho ^7+48877983 \rho
   ^6-590407188 \rho ^5-6662388180 \rho ^4
      \nonumber
    \\
    &&
   -10471753530 \rho ^3
   -3776755290 \rho ^2-82403160 \rho +17302320) \ln (\rho)
   \nonumber
    \\
    &&
   -\frac{1}{4762800 \rho }(559440 \rho ^9-25054275 \rho ^8+1507458405 \rho ^7-22969818054 \rho ^6
       \nonumber
    \\
    &&
   -93162469858 \rho ^5+29692805439 \rho
   ^4+78511998867 \rho ^3+6959893726 \rho ^2
    \nonumber
    \\
    &&
   -515509350 \rho +135660)
   -\frac{1}{396900 \rho ^2}(26460 \rho ^{10}-2980165 \rho^9-1295217 \rho ^8
    \nonumber
    \\
    &&
   -209585628 \rho ^7-1023663900 \rho ^6+147276360 \rho ^5+914828040 \rho ^4
    \nonumber
    \\
    &&
   +175288092 \rho ^3+304128 \rho^2-209475 \rho +11305) \ln (1-\rho )\,.
    \nonumber
    \\
    &&
\end{eqnarray}
The coefficients are also provided in the file ``Coef.m''.

\end{document}